\documentclass[aps,pra,twocolumn,preprintnumbers,amsmath,amssymb,floatfix,showpacs,superscriptaddress]{revtex4-2}

\usepackage{dcolumn}
\usepackage{xcolor} 
\usepackage[utf8x]{inputenc}
\usepackage{graphics}
\usepackage{epsfig}
\usepackage{subfigure}
\usepackage{rotating}
\usepackage{mathrsfs}
\usepackage{amsfonts}
\usepackage{amsmath}
\usepackage{amssymb}
\usepackage{placeins}
\usepackage{bm}
\usepackage{pgfplots}
\usepackage{soul}
\usepackage{fixmath}
\usepackage{physics}
\usepackage[colorlinks]{hyperref}

\usepackage{comment}

\newcommand{\transp}{\ensuremath{\mathrm{T}}}

\newtheorem{proposition}{Proposition}

\begin{document}

\title{Hidden quantum correlations in cavity-based quantum optics}

\author{Bakhao Dioum}
\affiliation{University of Lille, CNRS, UMR 8523 -- PhLAM -- Physique des Lasers Atomes et Mol\'{e}cules, F-59000 Lille, France}%
\email{bakhao.dioum@univ-lille.fr}
\author{Virginia D'Auria}%
\affiliation{Universit\'{e} C\^{o}te d’Azur, CNRS, Institut de physique de Nice, France}
\author{Giuseppe Patera}%
\affiliation{University of Lille, CNRS, UMR 8523 -- PhLAM -- Physique des Lasers Atomes et Mol\'{e}cules, F-59000 Lille, France}%

\date{\today}

\begin{abstract}
In multimode optical systems, the spectral covariance matrix encodes all the information about quantum correlations
between the quadratures of Gaussian states. 
Recent research has revealed that, in scenarios that are more common than previously thought, part of these correlations remain inaccessible to standard homodyne detection scheme. Formally, this effect can be attributed to a non-real spectral covariance matrix.
In this work, we provide a systematic framework and explicit criteria for identifying experimental configurations leading to such a behavior. This study will facilitate the proper exploitation and optimal engineering of CV quantum resources.
\end{abstract}

\maketitle

\section{\label{sec:I}Introduction}

Quantum states of light which exhibit nonclassicality in the continuous-variables (CV) regime~\cite{Braunstein2005a, Adesso2014a} are at the heart of many protocols for the processing of the quantum information such as quantum computation via cluster states~\cite{Menicucci2006, Gu2009}, quantum metrology~\cite{Giovannetti2004} and quantum communication~\cite{Ferraro2005,Weedbrook2012}. In particular Gaussian CV states have been effectively employed in quantum teleportation~\cite{Braunstein1998a, Furusawa1998, Braunstein2005a}, cryptography~\cite{Grosshans2002, Grosshans2003, Weedbrook2012}, and computation~\cite{Chen2013, Yokoyama2013, Menicucci2006, Adesso2014a} due to their mathematical simplicity with the existence of well-established theoretical tools to describe their properties~\cite{Weedbrook2012, Adesso2014a}. Single and multimode Gaussian states can be generated and manipulated experimentally in a variety of physical systems, ranging from light fields to atomic ensembles~\cite{Furasawa1998, Sherson2006}.
Nonclassicality in these states is generally observed through field quadrature measurements, with a key indicator being the variance of certain quadratures falling below the shot noise limit~\cite{Simon1994, VanLoock2003}. 
The second-order momenta of quadratures are collected in the covariance matrix that, for Gaussian states, up a displacement leading to zero mean values, faithfully replicate the density operator, thus offering a full characterization of these quantum states~\cite{Martinelli2023, Zhang2017, Brandao2022}. 

In cavity-based optical systems, boson field operators are slowly-varying with time in the interaction picture. To capture the temporal properties of these fields, second-order momenta are collected in a correlation matrix~\cite{Kolobov2011} that is function of two different times $t$ and $t'$. 

For stationary-in-time systems, the correlation matrix depends only on the time difference $t - t'$. For such systems, one can introduce its Fourier transform that depends only on one frequency $\omega$~\cite{Kolobov2011}. This object describes the frequency-dependent quantum correlations between the spectral components of amplitude and phase quadratures of single or multimode states. 
While in the literature this matrix is generally assumed to be real~\cite{Chembo2016, DAuria2009, Laurat2005, Guidry2023, Meng2020, Isaksen2023}, recent research reveals that scenarios leading to complex spectral covariances, 
signature of the presence of hidden squeezing ~\cite{Gouzien2023}, are more common than previously anticipated. 
Traditional measurement approaches, such as homodyne detection (HD)~\cite{Gu2009, Lvovsky2009a, Yurke1985},
while effective in characterizing real spectral covariance matrices ~\cite{Gouzien2020} and hence standard squeezing, encounter fundamental limitations when these are complex~\cite{Barbosa2013a,Barbosa2013b,Buchmann2016}. 
This indicates that a substantial portion of quantum correlations remains elusive to HD.
Such a limitation not only prevents accurate quantum state reconstruction, but also poses challenges in fully exploiting and manipulating these quantum states for quantum information processing~\cite{Lvovsky2020}.

Current literature points to the existence of these hidden correlations~\cite{Barbosa2013b,Gouzien2020, Gouzien2023} with some focus on more advanced detection methods such as  ``resonator detection"~\cite{Barbosa2013a,Barbosa2013b}, ``synodine detection"~\cite{Buchmann2016} or ``interferometers with memory effect"~\cite{Dioum2024}.
However, a systematic framework for precisely identifying the underlining interactions responsible for this behaviour is missing.

In this work, we provide precise criteria on the system's physical parameters (i.e.\ damping rates, mode hopping and pair production terms)  to determine whether a spectral covariance matrix is real or complex. These criteria allow predicting the nature of the Gaussian state solely from the system's interaction parameters without the necessity of fully characterizing the spectral covariance matrix or the complete quantum state reconstruction. As a consequence, our results allow the appropriate choice of the measurement strategies for a better characterization and exploitation of these quantum states.

For example, in their study of broadband squeezing generation in microresonators, Vaidya et al.~\cite{Vaidya2020a} relied on HD for the full characterization of the generated broadband squeezing. Using a configuration of parameters similar to their experimental setup, our criteria predict the absence of hidden squeezing. This confirms that HD was indeed sufficient for the measurement of the optimal squeezing produced by their device.
On the other side, in the case of the generation of broadband squeezing in dual-pumped microresonators~\cite{Seifoory2022}, the authors relying solely on HD schemes, attributed the lower detected squeezing levels entirely to ``parasitic" effects from additional interactions. However, our criteria predict the presence of hidden correlations. This indicates that while parasitic effects do play a role, more squeezing would be available than what is detected by HD. 
This is because, as we showed in~\cite{Dioum2024},
these interactions affect the time/frequency modes of the squeezing in a way that HD can never reach optimal mode-matching, thus, always resulting into a sub-optimal measurement of squeezing.
In a complementary way, this criteria also provides a guideline for the design of optical devices to tailor or avoid these complex correlations.

We first present, in Section \ref{sec:II}, the theoretical formalism for spectral covariance matrices in multimode quantum systems and apply it to the cavity-based quantum systems' framework. 
We derive, in Section \ref{sec:III}, explicit conditions on damping and interaction matrices that elucidate the transitional boundaries between real and complex spectral covariance. 
We show that beyond trivial interactions, mode-dependent damping rates and a non-symmetric product between the different types of interactions lead to complex quantum correlations.
To illustrate our predictions and build intuition for their application, we present some case studies of accessible few-mode systems in Section \ref{sec:IV}. The simplicity of these examples is useful for understanding the theory and gaining an intuitive grasp of the multimode quantum effects embodied in complex covariance matrices.
We show why single-mode systems inherently exhibit standard squeezing. 
We demonstrate how introducing asymmetry, either through mode-dependent dissipation and detuning or nonlinear interactions, give rise to hidden squeezing. Finally, we present a specific case where our criteria reveal that, beyond the previously assumed parasitic effects, there are still fundamental limitations to homodyne detection in capturing the full squeezing properties of certain quantum states.

\section{\label{sec:II}Theoretical Framework}

\subsection{\label{sec:IIA} Complex spectral covariance matrix }
We consider $N$ boson modes with annihilation and creation operators $\hat{a}_m(t)$ and $\hat{a}^{\dagger}_m(t)$, for $m=\left\{1,\cdots,N \right\}$, 
which are slowly-varying in time (in the interaction picture)  and satisfy the commutation relations $[\hat{a}_m(t),\hat{a}^{\dagger}_n(t')]=\delta_{mn}\delta(t-t')$.
In the quadrature representation, we define the column vector $\hat{\mathbf{R}}(t)=\left(\hat{\bm{x}}(t)|\hat{\bm{y}}(t)\right)^{\transp}$
where $\hat{\bm{x}}(t)=\left(\hat{x}_1(t),\cdots,\hat{x}_N(t)\right)^{\transp}$ and
$\hat{\bm{y}}(t)=\left(\hat{y}_1(t),\cdots,\hat{y}_N(t)\right)^{\transp}$.
The quadratures $\hat{x}_m(t)=\left(\hat{a}_m(t)+\hat{a}^{\dagger}_m(t)\right)/\sqrt{2}$ and $\hat{y}_m(t)= i\left(\hat{a}^{\dagger}_m(t)-\hat{a}_m(t)\right)/\sqrt{2}$ 
are hermitian operators, corresponding to the amplitude and phase observables, respectively.

Up to a displacement leading to zero mean values, the statistical properties of Gaussian states are fully captured by the covariance matrix:
\begin{align} 
\sigma(t,t')&=\frac{1}{2}\left\langle \hat{\mathbf{R}}(t) \hat{\mathbf{R}}^{\transp}(t')+\left(\hat{\mathbf{R}}(t')\hat{\mathbf{R}}^{\transp}(t)\right)^{\transp}\right\rangle,
\label{eq:covar_temp}
\end{align}
that must be symmetric under the exchange $(m,t) \rightleftharpoons (n,t')$ for it to represent a physical state.
A similar object is better known in the literature as ``correlation matrix"~\cite{Kolobov2011,Martinelli2023} even if in some cases only the first term in Eq.~\eqref{eq:covar_temp} appears.

For stationary processes, the covariance matrix depends only on $\tau =t'-t$, and
it is possible to define the spectral covariance matrix as the Fourier transform of~\eqref{eq:covar_temp}:
\begin{align}
\sigma(\omega)&=
\frac{1}{2}\left\langle \hat{\mathbf{R}}(\omega) \hat{\mathbf{R}}^{\transp}(-\omega)+\left(\hat{\mathbf{R}}(-\omega)\hat{\mathbf{R}}^{\transp}(\omega)\right)^{\transp}\right\rangle.
\label{eq:covar_spect}   
\end{align}

In the literature, this quantity can be referred to as ``hermitian covariance matrix"~\cite{Buchmann2016} or ``spectral density matrix"~\cite{Barbosa2013a, Barbosa2013b,Martinelli2023}, even if in some cases, the definition does not take into account the second term in Eq.~\eqref{eq:covar_spect}.

The spectral quadrature operators $\hat{\mathbf{R}}(\omega)$, given by the symmetric Fourier transform of their temporal counterparts
\begin{align}
\hat{\mathbf{R}}(\omega)& = \int \frac{dt}{\sqrt{2\pi}} \mathrm{e}^{-i \omega t} \hat{\mathbf{R}}(t),
\label{eq:fourier_R}
\end{align}
have commutators $[\hat{\mathbf{R}} (\omega),\hat{\mathbf{R}}^{\transp} (\omega')]= i\Omega\delta(\omega+\omega')$ with $\Omega$ the symplectic form~\cite{SympForm} and verify the relation 
$\hat{\mathbf{R}}^{\dagger}(\omega)
=\hat{\mathbf{R}}(-\omega)$ in order to correspond to \textit{bona fide} quadrature operators in time domain. However, these spectral quadratures do not correspond to single-mode spectral quadratures. This can be seen from the individual mode quadratures $\hat{x}_m(\omega)=\left(\hat{a}_m(\omega)+[\hat{a}_m(-\omega)]^\dagger\right)/\sqrt{2}$ and $\hat{y}_m(\omega)= i\left([\hat{a}_m(-\omega)]^\dagger-\hat{a}_m(\omega)\right)/\sqrt{2}$, which show that they are connected to a linear combination of upper ($\omega$) and lower ($-\omega$) sidebands with respect to the central frequencies~\cite{Martinelli2023, Barbosa2013b,Lvovsky2016}.

As a consequence of the properties of these spectral quadrature operators, the spectral covariance matrix can in general be complex-valued, as far as it is conjugate-symmetric $\sigma^{\ast}(\omega)=\sigma(-\omega)$. This also means that, with respect to the exchange $\omega\leftrightarrow -\omega$, its real part must be symmetric $\sigma_\mathrm{R}(\omega)=\sigma_\mathrm{R}(-\omega)$ and its imaginary part must be antisymmetric $\sigma_{\mathrm{I}}(\omega)=-\sigma_{\mathrm{I}}(-\omega)$.

Therefore, sideband-symmetric quantum states necessarily have a real spectral covariance matrix: physically this means that the sideband modes symmetric (with respect to the carrier at frequencies $\pm \omega$) are in the same quantum state. In this condition, an effective single-mode description can be applied, and complete characterization can be obtained through standard homodyne detection~\cite{Lvovsky2016}. In contrast, for quantum states that are not symmetric, the imaginary part $\sigma_{\mathrm{I}}(\omega)$ is not null.
In this case, symmetric sideband modes are not in the same quantum state and homodyne detection is not able to provide a full characterization~\cite{Barbosa2013b,Barbosa2013b,Buchmann2016}. States with such behaviour have been identified in two-mode optical parametric oscillators~\cite{Barbosa2018a}, in optomechanical systems~\cite{Buchmann2016}, in microring resonators~\cite{Gouzien2023} and, more generally, in any multimode system whose dissipative nonlinear dynamics is reduced to an effective quadratic Hamiltonian after the linearization around stable stationary solutions.
The proper detection of these ``hidden" or ``complex" correlations rather requires more sophisticated schemes such as ``resonator detection"~\cite{Barbosa2013b,Barbosa2013b}, ``synodine detection"~\cite{Buchmann2016} or ``interferometers with memory effect"~\cite{Dioum2024}.

\subsection{\label{sec:IIB} Cavity-based multimode quantum optics}
When the nonlinear dynamics of a system of $N$ boson modes occur in a cavity, it can be linearized around a stable classical steady-state solution. Thus, in the most general case, it can be associated to the following effective quadratic Hamiltonian:
\begin{equation}
\hat{H} = \hbar \sum_{m,n}G_{m,n} \hat{a}_{m}^{\dagger}\hat{a}_{n}+ \frac{\hbar}{2}\sum_{m,n}[F_{m,n} \hat{a}_{m}^{\dagger}\hat{a}^{\dagger}_{n} + H.c.],
\label{eq:hamiltonian}
\end{equation}
where $F$ and $G$ are $N \times N$ complex matrices with $F$ being symmetric ($F=F^{\transp}$), describing pair-production processes 
and $G$ being Hermitian ($G= G^{\dagger}$), describing mode-hopping processes.

The corresponding set of linear quantum Langevin equations for the quadratures operators is given by
\begin{align}
\frac{\mathrm{d}\hat{\mathbf{R}}(t)}{\mathrm{d}t}&=
(-\Gamma+\mathcal{M})\hat{\mathbf{R}}(t)+
\sqrt{2\Gamma}\,\hat{\mathbf{R}}_{\mathrm{in}}(t)
\label{eq:quantum_lang_r}
\end{align}
where $\Gamma=\mathrm{diag}\{\gamma_1,\ldots,\gamma_N|\gamma_1,\ldots,\gamma_N\}$ is a $2N\times 2N$ diagonal matrix of the damping rates that in general can be mode-dependent and that can account for multiple sources of losses or couplings with the
external environment and $\hat{\mathbf{R}}_{\mathrm{in}}$ is the column vector collecting the quadratures of the input state. The  $2N\times 2N$ mode interaction matrix $\mathcal{M}$, which writes
\begin{equation}
\mathcal{M}=
\left(
\begin{array}{c|c}
\mathrm{Im}\left[G+F\right] & \mathrm{Re}\left[G-F\right]
\\
\hline
-\mathrm{Re}\left[G+F\right] & -\mathrm{Im}\left[G+F\right]^\transp
\end{array}
\right)
\label{eq:M_cali}
\end{equation}
encapsulates all mode interactions.
The goal of the next sections is to elucidate the relation between $\mathcal{M}$ and spectral covariance matrix and to establish conditions for distinguishing classes of quantum states with detectable or hidden correlations.

The quadratures $\hat{\mathbf{R}}_{\mathrm{out}}(t)$ of the system's output can be obtained through the input-output relations $\hat{\mathbf{R}}_{\mathrm{in}}(t)+\hat{\mathbf{R}}_{\mathrm{out}}(t)=\sqrt{2\Gamma}\,\hat{\mathbf{R}}(t)$~\cite{gardiner2004}.
In the frequency domain, they are expressed in terms of the input quadratures through the transfer function matrix $S(\omega)$
\begin{equation}
\hat{\mathbf{R}}_{\mathrm{out}}(\omega)= S(\omega)\hat{\mathbf{R}}_{\mathrm{in}}(\omega), 
\end{equation}
where $S(\omega)$ is the matrix-valued function
\begin{equation}
S(\omega)=\sqrt{2\Gamma}\left(i\omega\mathbb{I}+\Gamma-\mathcal{M}\right)^{-1}\sqrt{2\Gamma}-\mathbb{I}.
\label{eq:transfer_funct}
\end{equation}
Here, $\mathbb{I}$ is the $2N \times 2N$ identity matrix. To have \textit{bona fide} quadratures in time domain, $S(\omega)$ must be conjugate-symmetric, $S^{\ast}(\omega)=S(-\omega)$, and ``$\omega$-symplectic'' that stands for a smooth (analytic) matrix-valued function that is conjugate-symplectic~\footnote{The conjugate-symplectic group Sp$^*(2N,\mathbb{C})$ is defined as the set of $2N\times 2N$ complex matrices such that $S\Omega S^\dagger$, with $\Omega$ the symplectic form.}. The first condition is directly verified by inspecting Eq.~\eqref{eq:transfer_funct}. The second is guaranteed by the fact that $\mathcal{M}$ is a Hamiltonian matrix (\textit{i.e.}\ $(\Omega\mathcal{M})^\transp=\Omega\mathcal{M}$)~\cite{Gouzien2020}.

Assuming vacuum input state and injecting Eq.~\eqref{eq:transfer_funct} in Eq.~\eqref{eq:covar_spect}, the covariance matrix for the output quadratures is
\begin{align}
\sigma_{\mathrm{out}}(\omega)&=\frac{1}{2\sqrt{2\pi}}
S(\omega)S^{\transp}(-\omega).
\label{eq:covar_spect_2}
\end{align}
To gain further insights into relating $S(\omega)$ to the output spectral covariance matrix, it is useful to explicitly decompose it into is real symmetric and imaginary antisymmetric parts $S(\omega) = S_{\mathrm{R}}(\omega) + iS_{\mathrm{I}}(\omega)$ 
defined as
\begin{align}
S_{\mathrm{R}}(\omega)&=\sqrt{2\Gamma}\Big[
(\Gamma-\mathcal{M})
\mathrm{Inv}(\omega)
\Big]\sqrt{2\Gamma}-\mathbb{I}
\label{eq:transf_funct_real}
\\
S_{\mathrm{I}}(\omega)&=\sqrt{2\Gamma}\Big[
-\omega
\mathrm{Inv}(\omega)
\Big]\sqrt{2\Gamma},
\label{eq:transf_funct_imag}
\end{align}
where $\mathrm{Inv}(\omega)$ is defined as $\mathrm{Inv}(\omega) = [\omega^{2}\mathbb{I} +(\Gamma-\mathcal{M})^{2}]^{-1}$. We can further factor the expression of $S_{\mathrm{R}}(\omega)$ in Eq.~\eqref{eq:transf_funct_real} as in Appendix~\ref{ap:II} (see Eq.~\eqref{seq:s_r_2} for more details). 
The real part of $S_{\mathrm{R}}(\omega)$ therefore can be rewritten as:
\begin{align}
S_{\mathrm{R}}(\omega)&=\frac{1}{\sqrt{2\Gamma}}\Big[
\Big(\Gamma^2+[\mathcal{M}, \Gamma]-\mathcal{M}^2- \omega^2 \mathbb{I}\Big)
\mathrm{Inv}^{-1}(\omega)
\Big]\sqrt{2\Gamma},
\label{eq:transf_funct_real_2}
\end{align}

where $[\mathcal{M}, \Gamma]= \mathcal{M}\Gamma-\Gamma \mathcal{M}$.

Using Eq.~\eqref{eq:covar_spect_2}, we decompose the spectral covariance matrix into its real and imaginary parts: 
$\sigma_{\mathrm{out}}(\omega) = \sigma_\mathrm{R}(\omega) + i \sigma_{\mathrm{I}}(\omega)$  where
\begin{align}
\sigma_{\mathrm{R}}(\omega)
&= \frac{1}{2\sqrt{2\pi}}\Big(
S_{\mathrm{R}}(\omega)S_{\mathrm{R}}^{\transp}(\omega) +
S_{\mathrm{I}}(\omega)S_{\mathrm{I}}^{\transp}(\omega)\Big),
\label{eq:covar_spect_split_2}
\\
\sigma_{\mathrm{I}}(\omega)
&=\frac{1}{2\sqrt{2\pi}}\Big(
S_{\mathrm{R}}(\omega)S_{\mathrm{I}}^{\transp}(\omega) -
S_{\mathrm{I}}(\omega)S_{\mathrm{R}}^{\transp}(\omega)\Big).
\label{eq:covar_spect_split}
\end{align}
Using Eqs.~\eqref{eq:covar_spect_split},~\eqref{eq:covar_spect_split_2}, we can relate the reality or complexity of $\sigma_{\mathrm{out}}(\omega)$ to conditions on the real and imaginary parts of $S(\omega)$.

\section{\label{sec:III}Boundaries to complex spectral covariance matrices}
\begin{figure}[ht!]
\centering
\includegraphics[width=1\columnwidth]{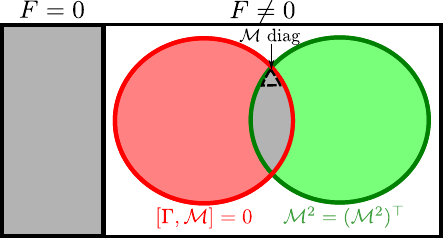}
\caption{Conditions for a real-valued spectral covariance matrix $\sigma_{\mathrm{out}}(\omega)$ are represented by the grey areas. The left region represents the trivial case where pair-production processes are absent ($F = 0$), which always leading to a real $\sigma_{\mathrm{out}}(\omega)$. The right region depicts the scenario where pair-production processes are present ($F \neq 0$). In this case, the conditions for achieving a real-valued $\sigma_{\mathrm{out}}(\omega)$ are represented by the overlap of the red circle, corresponding to the commutation between the damping and mode interaction matrices $[\Gamma, \mathcal{M}] = 0$, and the green circle, corresponding to the symmetry of the mode interaction matrix squared $\mathcal{M}^2 = (\mathcal{M}^2)^T$. This overlap region (gray) delineates the boundaries between real and complex $\sigma_{\mathrm{out}.}(\omega)$. Within this region, we represented a subset ``$\mathcal{M}$ diag" corresponding to the case where $\mathcal{M}$ is diagonal. Here, both conditions necessary for a real spectral covariance matrix, are verified.}
\label{fig:conditions}
\end{figure}
In this section, we now seek conditions on the damping matrix $\Gamma$ and the mode interaction matrix $\mathcal{M}$ that lead to a complex-valued or real-valued spectral covariance matrix $\sigma_{\mathrm{out}}(\omega)$.
From Eq.~\eqref{eq:transf_funct_imag}, it is evident that $S(\omega)$ and consequently $\sigma_{\mathrm{out}}(\omega)$ are always real at $\omega=0$.

A trivial case, where a complex $S_{\mathrm{out}}(\omega)$ leads to a real $\sigma_{\mathrm{out}}(\omega)$, occurs in the absence of pair-production processes, i.e.\ $F =0$. These processes are, for example, responsible for the generation of multimode entanglement and squeezing in optical parametric oscillators~\cite{Fabre2020, Patera2012, Arzani2018}. Without pair-production processes, the transfer function in Eq.~\eqref{eq:transfer_funct} becomes unitary, yielding a $\sigma_{\mathrm{out}}(\omega)$ proportional to the identity matrix and therefore real, as illustrated in the gray region of Fig.~\ref{fig:conditions}. However, this case is not particularly interesting when the primary objective of quantum optical systems is the generation of non-classical quantum states.

When $F \neq 0$, the structure of Eq.~\eqref{eq:covar_spect_split} suggests that a necessary and sufficient condition for the output spectral covariance to be real is that the product $S_{\mathrm{R}}(\omega) S_{\mathrm{I}}^{\transp}(\omega)$ must be symmetric. By substituting Eq.~\eqref{eq:transf_funct_real} and \eqref{eq:transf_funct_imag} into this product, we obtain:
\begin{align}
S_{\mathrm{R}}(\omega)S_{\mathrm{I}}^{\transp}(\omega)&= 
\frac{-2\omega}{\sqrt{2\Gamma}}
A(\omega)B(\omega) \sqrt{2\Gamma},
\label{eq:product}
\end{align}

where $A(\omega)=\Gamma^2+[\mathcal{M}, \Gamma]-\mathcal{M}^2-\omega^2 \mathbb{I}$ and $B(\omega) = \mathrm{Inv}(\omega) \Gamma \mathrm{Inv}^{\transp}(\omega)$ is a symmetric matrix.  By examining the difference $S_{\mathrm{R}}(\omega)S_{\mathrm{I}}^{\transp}(\omega)-S_{\mathrm{I}}(\omega)S_{\mathrm{R}}^{\transp}(\omega)$ (See Eq.~\ref{seq:difference} in Appendix~\ref{ap:II}), it becomes evident that the product $S_{\mathrm{R}}(\omega)S_{\mathrm{I}}^{\transp}(\omega)$ is symmetric when $A(\omega)B(\omega)\Gamma = \Gamma B(\omega) A^{\transp}(\omega)$. This condition holds only when these 2 following conditions are met at the same time:
\begin{itemize}
    \item $\Gamma$, $A(\omega)$ and $B(\omega)$ commute pair-wise,
    \item $A(\omega)$ is symmetric : $A^{\transp}(\omega) = A(\omega)$.
\end{itemize}

We show in Appendix \ref{ap:III} that the first condition is equivalent to $\Gamma$ and $\mathcal{M}$ commuting $[\Gamma, \mathcal{M}]=0$. For the second condition, $A(\omega)$ is symmetric if and only if the following two requirements are met:
(i) $[\Gamma,\mathcal{M}] = [\Gamma,\mathcal{M}^{\transp}]$ and
(ii) $\mathcal{M}^2 = (\mathcal{M}^2)^{\transp}$.
When $\Gamma$ and $\mathcal{M}$ commute, condition (i) is automatically satisfied, leaving only condition (ii). This leads us to the following proposition:
\begin{proposition}\label{prop cov_real}
For $F\neq 0$, necessary and sufficient conditions for the spectral covariance to be real are $[\Gamma,\mathcal{M}]=0$ and $\mathcal{M}^{2}$ symmetric.
\end{proposition}  
This proposition is visually represented in the right side of Fig.~\ref{fig:conditions}, where the gray overlap region of the Venn diagram depicts the intersection of these two conditions: the red subset corresponds to the commutation between the damping and mode interaction matrices, $[\Gamma, \mathcal{M}] = 0$, while the green subset corresponds to the symmetry of the mode interaction matrix squared, $\mathcal{M}^2 = (\mathcal{M}^2)^T$. Failing to satisfy either of these conditions necessarily leads to an asymmetry of the product $S_{\mathrm{R}}(\omega)S_{\mathrm{I}}^{\transp}(\omega)$ and to a complex spectral covariance matrix.

On one hand, since $\Gamma$ is diagonal, the commutator $[\Gamma, \mathcal{M}]$ can be zero only if either (i) $\Gamma$ is proportional to the identity matrix, corresponding to mode-independent damping rates, or (ii) $\mathcal{M}$ is diagonal, or both. It is straightforward to see that the scenario (ii) leads, to a configuration where $G = 0$ and $F = iD$, with $D$ a real diagonal matrix corresponding to the case of $N$ decoupled squeezed modes. Since a diagonal $\mathcal{M}$ also implies $\mathcal{M}^{2}$ symmetric, this situation directly leads to a real spectral covariance matrix. This is represented in Fig.~\ref{fig:conditions} by a dashed region in the intersection between the red and green subsets.

On the other hand, to see how the symmetry of $\mathcal{M}^{2}$ reflects on the structure of $G$ and $F$, it is better to move to the representation of complex amplitudes, where $\mathcal{M}$ transforms into the 
$2N\times 2N$ matrix $M=i\, L^{\dagger}\mathcal{M}L$:
\begin{equation}
M=
\left(
\begin{array}{c|c}
G & F
\\
\hline
-F^{\ast} & -G^{\ast}
\end{array}
\right),
\label{eq:M_droit}
\end{equation}
where $L$ is the unitary matrix performing a change of basis from the boson operators $\{\hat{a}_k,\hat{a}^\dagger_k\}$ to the quadratures operators $\{\hat{x}_k,\hat{y}_k\}$ (for $k=1,\ldots,N$):
\begin{equation}
L = \frac{1}{\sqrt{2}}\left(\begin{array}{c|c}I_{N} & I_{N}\\ \hline-\mathrm{i}I_{N} & \mathrm{i}I_{N}\end{array}\right)
\end{equation}
and $I_{N}$ is the identity matrix of dimension $N\times N$.
The symmetry of $\mathcal{M}^2$ is equivalent to the hermiticity of $M^2$ where
\begin{align}
M^{2}
= \begin{pmatrix}
G^{2}-FF^{\ast} & GF-FG^{\ast} \\
-(GF-FG^{\ast})^{\dagger} & (G^{2}-FF^{\ast})^{\transp}
\end{pmatrix}.
\label{eq:M^2}
\end{align}
By examining its block structure, the hermiticity of $M^2$ implies the hermiticity of its block elements. Since $G=G^\dagger$ and $F=F^\transp$, the diagonal blocks are always hermitian. However, off-diagonal blocks are hermitian only when $GF-FG^{\ast} = 0$, or equivalently $GF = (GF)^{\transp}$. One can easily verify that this condition is satisfied, for example, in the case where a real-valued $G$ commute with $F$ as we will show in the practical examples in Section \ref{sec:IV}. This condition leads us to the following proposition, equivalent to proposition~\ref{prop cov_real}:
\begin{proposition}\label{prop GF}
For $F\neq 0$, necessary and sufficient conditions for the spectral covariance to be real are $[\Gamma,\mathcal{M}]=0$ and that the product $GF$ is symmetric $GF = (GF)^{\transp}$.
\end{proposition}

Propositions~\ref{prop cov_real} and \ref{prop GF} elucidate the transitional boundary between standard and hidden squeezing. Notice that, beyond the trivial cases $F=0$ and $\mathcal{M}$ diagonal, a generic $M$, for which $GF = (GF)^{\transp}$, leads to a real spectral covariance matrix only in the case of mode-independent losses, i.e.\ when $\Gamma$ is proportional to the identity matrix.

These conditions provide insights into the underlying interactions that govern the system's dynamics and can guide the engineering of complex multimode correlations.

\section{\label{sec:IV} Case studies}

To illustrate the developed framework, we investigate several case studies of cavity-based optical quantum systems. 
We focus on low-dimensional examples, ranging from single- to three-mode systems, where the analysis remains tractable while clearly illustrating the key concepts.
In particular, we demonstrate how introducing either  mode-dependent damping rate or asymmetry in the product of $F$ and $G$, leads 
to complex spectral covariance matrices. 
Since our examples involve cavity longitudinal modes, the mode-hopping processes included in $G$ describe processes such as dispersion induced mode-dependent detunings from resonance, self-phase and cross-phase modulations. Pair-production processes in $F$ instead accounts for parametric amplification processes~\cite{Gouzien2023}. 

For all considered examples, we compare the optimal squeezing spectrum with squeezing measured via a standard homodyne measurement, to show that in the case of a complex spectral covariance matrix, part of the squeezing is hidden to HD. Optimal squeezing is obtained via analytic Bloch-Messiah decomposition (ABMD)~\cite{Gouzien2020} and implemented in Python as an adaptation to conjugate symplectic matrices of the smooth singular value decomposition~\cite{Dieci2024,Pugliese_matlab}.

\subsection{\label{subsec:IVA} Single-mode detuned OPO}

As the most minimal system, we first analyse a single-mode detuned degenerate optical parametric oscillator (OPO)~\cite{Fabre1990, Fabre1989}. 
This system consists of a nonlinear cavity of damping rate $\gamma$, supporting a single resonant mode at frequency $\omega_0$.  
Driving the cavity with a pump laser at frequency $\omega_p$ leads to parametric downconversion~\cite{Christ2013c} of coupling strength $g$,
coupling pump photons to degenerate pairs of signal photons at frequency $\omega_s$, close to the cavity resonance $\omega_0$. The signal frequency $\omega_s$ is slightly detuned from resonance by 
$\Delta=\omega_s-\omega_0$.
Below threshold, the dynamics of the signal field can be described, in the interaction picture, by the quantum Langevin equations in the form of Eq.~\eqref{eq:quantum_lang_r} in the quadrature basis $\hat{\bf{R}}=(\hat{x}|\hat{y})^{\transp}$.

The corresponding mode interaction matrix of the system is described by $G = \Delta$ and $F=ig$~\cite{Gouzien2020}, which in this case are scalar values.
These interaction elements allow to immediately predict that the single-mode detuned OPO cannot exhibit complex spectral covariance. 
First, without multiple modes, there cannot be any relative asymmetry between dissipation rates contained in $\Gamma$. 
Then, since there is only one mode, the matrices $F$ and $G$ are simply reduced to scalars, which guarantees their product $GF$ is always symmetric, 
satisfying both conditions for a real covariance matrix in Proposition~\ref{prop GF}.
Intuitively, there are no possibilities to break these symmetries within a single-mode system. 
Consequently, the spectral covariance matrix $\sigma_{\mathrm{out}}(\omega)$ of the single mode degenerate OPO must be real-valued, despite a complex transfer function $S_{\mathrm{out}}(\omega)$.

\begin{figure}[t!]
\centering
\includegraphics[width=1\columnwidth]{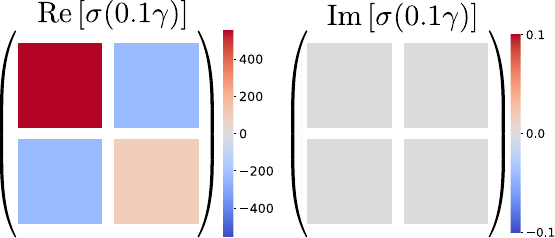}
\caption{Real (left) and imaginary (right) part of the elements of the $2\times2$ output spectral covariance matrix $\sigma_{\mathrm{out}}(\omega)$. They are shown for a detuned degenerate single mode OPO of parameters  $\Delta = \gamma$, $g = 1.38 \gamma$ and $\omega= 0.1\gamma$. For this choice of parameters $(GF)^\transp= GF$.
We see the spectral covariance matrix presents no imaginary elements, which means it is real-valued.}
\label{fig:single_mode_covar}
\end{figure}
%
\begin{figure}[t!]
\centering
\includegraphics[width=1\columnwidth]{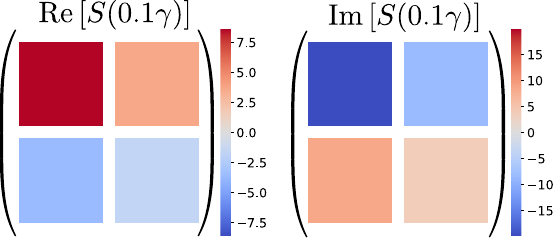}
\caption{Real (left) and imaginary part (right) of the elements of the $2\times2$ transfer function $S_{\mathrm{out}}(\omega)$ of a detuned single mode OPO for $\Delta = \gamma$, $g=1.38\gamma$ and $\omega= 0.1\gamma$. Despite a corresponding real spectral covariance $\sigma_{\mathrm{out}}(\omega)$ (Fig.~\ref{fig:single_mode_covar}), $S_{\mathrm{out}}(\omega)$ is complex}
\label{fig:single_mode_transfer}
\end{figure}
\begin{figure}[t!]
\centering
\includegraphics[width=1\columnwidth]{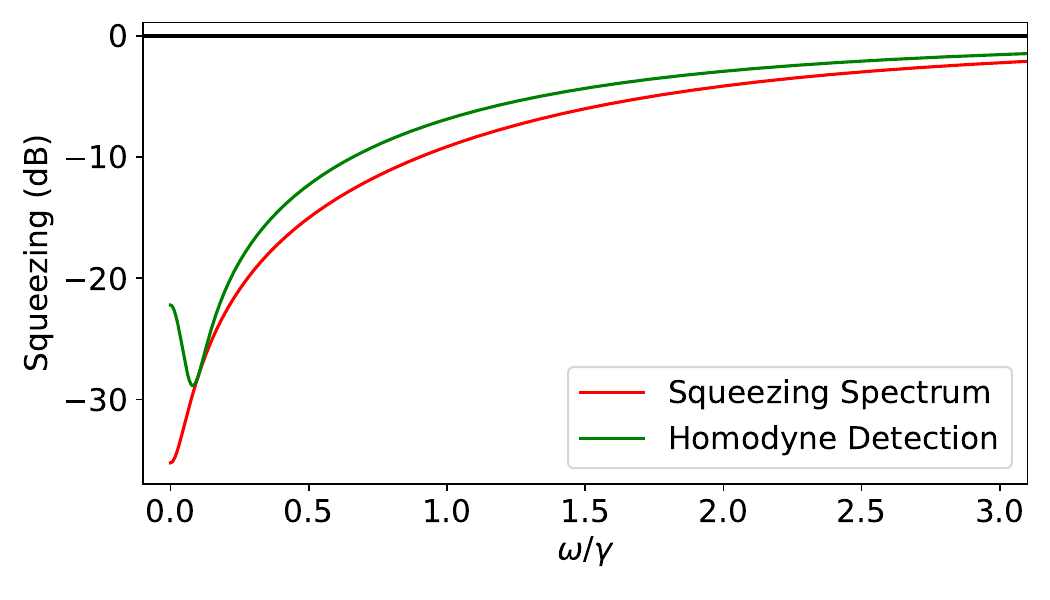}
\caption{Comparison between the squeezing spectrum (red) for a single-mode OPO and the spectrum measured via standard homodyne detection (HD) (green) optimized at $\omega = 0.1\gamma$. The corresponding real spectral covariance matrix is depicted in Fig.~\ref{fig:single_mode_covar}. At this particular frequency, the two curves coincide, indicating that the optimal squeezing level is fully accessible through standard HD.}
\label{fig:single_mode_HD}
\end{figure}

For example, for  given parameters of the single mode OPO ($\Delta=\gamma, g=1.38\gamma$), we show in Fig.~\ref{fig:single_mode_covar} and Fig.~\ref{fig:single_mode_transfer}, 
the four elements of the $2\times 2$ spectral covariance matrix and transfer function respectively, at a representative nonzero frequency $\omega = 0.1 \gamma$. These plots are representative of the behavior of the spectral covariance and transfer function matrices at other nonzero frequencies.
In each figure, we plot on the left  the real parts of these elements and on the right their imaginary parts.  
We see from Fig.~\ref{fig:single_mode_covar} that the spectral covariance matrix presents no imaginary elements, 
despite a complex transfer function (see Fig.~\ref{fig:single_mode_transfer}).

The reality of the spectral covariance matrix ensures that optimal squeezing levels can be measured using a standard HD. In Fig.~\ref{fig:single_mode_HD}, we compare the optimal squeezing that is actually produced by the OPO (solid-red curve) with the squeezing measured by a standard HD scheme (solid-green curve). The phase of the local oscillator is chosen to optimize the squeezing detection at a given frequency (here $\omega=0.1\gamma$). The fact that the green curve touches the red at the selected frequency indicates that all the squeezing is detectable by the HD. A full reconstruction of the optimal squeezing spectrum can be realized by scanning the LO phase.

\subsection{\label{subsec:IVB} Two-mode optomechanical system}

To incrementally go beyond single-mode constraints, we consider a two-mode optomechanical cavity ~\cite{Mancini1994a, Fabre1994, Aspelmeyer2014}. Typically, this system consists of a driven high-finesse cavity mode with resonance frequency $\omega_0$ coupled to one mechanical deformation mode of one mirror of the cavity with frequency $\omega_m$.
A key feature of interest of this system is that the optical and mechanical modes exhibit different damping rates $\gamma$ and $\gamma_m$, 
corresponding to their coupling to independent noise reservoirs.
This conceptually simple asymmetry provides, as predicted by our criteria, the transition from real to complex spectral covariance matrix.
When the cavity is driven by a laser at frequency $\omega_p$, detuned from the resonance frequency $\omega_0$ by $\Delta= \omega_p - \omega_0$, 
photons exert a radiation pressure force on the mechanical resonator via their momentum transfer upon reflection. 
In the two-mode interaction picture, this couples the optical cavity intensity (photon states) to the mechanical displacement (phonon states) at a rate $g$ ~\cite{Mancini1994a, Fabre1994, Aspelmeyer2014}.

After linearization around a stable solution for a coherent pump, the system is described in the pump frequency rotating frame by a Hamiltonian in the form of 
Eq.~\eqref{eq:quantum_lang_r} and mode interaction matrices $G$ and $F$ described as follows:

\begin{align}
G = \begin{pmatrix}
\Delta & g \\
g & \omega_m
\end{pmatrix}, \quad
F = \begin{pmatrix}
0 & 2g \\
2g & 0
\end{pmatrix}.
\label{eq:opto_mech}
\end{align}

With no loss of generality, we consider the case where the coupling strength $g$ is real and the optical detuning $\Delta$ matches the mechanical frequency $\omega_m$. Under these conditions $G$ is real and commutes with $F$; combined with the symmetric property of $F$, this yields a symmetric $GF$ product. Nevertheless, for optomechanical systems, the time scale of the evolution of the optical field and that of the mechanical deformation of the mirror are, in general, very different. Therefore the difference in damping rates makes it impossible to satisfy Proposition 1 and leads to a complex spectral covariance matrix.

\begin{figure}[t!]
\centering
\includegraphics[width=1\columnwidth]{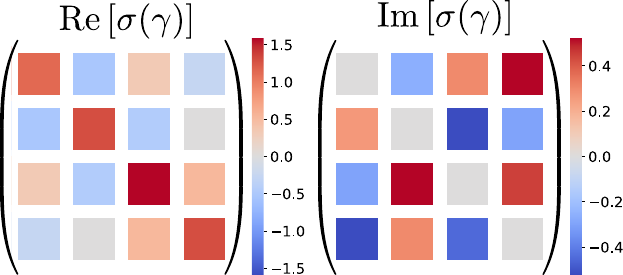}
\caption{ Spectral covariance matrix $\sigma_{\mathrm{out}}(\omega)$ for a two-mode optomechanical system with optical and mechanical damping rates respectively $\gamma$ and $\gamma_m = 0.001 \, \gamma$. For this choice of parameters $(GF)^\transp\neq GF$.
$\sigma_{\mathrm{out}}(\omega)$, at a given frequency $\omega=\gamma$, is $4\times4$ matrix where we have plotted the real (left) and 
imaginary part (right) of its 16 elements. We see the matrix exhibits nonzero imaginary components, indicating complex covariance arising solely from 
the mode-dependent damping rates. Other chosen numerical parameters are: $\Delta=\omega_m = \gamma$ and $g = 0.01\gamma$.}
\label{fig:opto_mec_covar}
\end{figure}
\begin{figure}[t!]
\centering
\includegraphics[width=1\columnwidth]{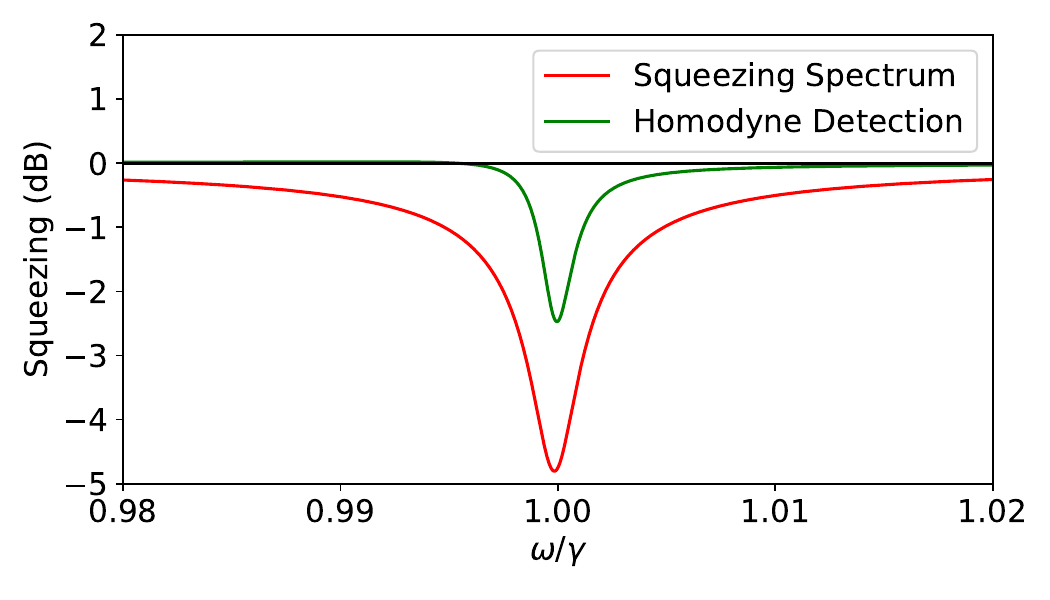}
\caption{Squeezing spectrum (red) for the two-mode optomechanical system whose complex spectral covariance matrix is depicted in Fig.~\ref{fig:opto_mec_covar} at $\omega = \gamma$. We represent the maximum achievable hybrid squeezing in red and a measure through standard homodyne detection on the hybrid mechanical-optical modes at that frequency. Unlike the single-mode case, the two curves do not coincide at that frequency, due to the complex nature of the spectral covariance matrix. The red curve corresponds to the maximum achievable squeezing, which is inaccessible through standard homodyne detection at nonzero frequencies ~\cite{Gouzien2023, Dioum2024}.}
\label{fig:opto_mec_HD}
\end{figure}

This is evident from Fig.~\ref{fig:opto_mec_covar} where we have plotted, at the frequency $\omega=\gamma$, the real (on the left) and imaginary (on the right) parts of the 16 elements contained in the $4\times4$ spectral covariance $\sigma_{\mathrm{out}}(\omega)$. We considered an optomechanical model with parameters : $\gamma_m = 0.001 \gamma$, $\Delta= \omega_m=\gamma$ and $g = 0.01\gamma$. The nonzero imaginary components, shown in Fig.~\ref{fig:opto_mec_covar}, 
illustrate that asymmetry in the decay rates alone is sufficient to generate complex correlations, characteristic of hidden squeezing ~\cite{Gouzien2023}. 
 
This is in line with our predictive results in Proposition~\ref{prop GF} and holds for all nonzero frequencies. 

In Fig \ref{fig:opto_mec_covar}, the red curve represents the maximum achievable squeezing in the hybrid mechanical-optical mode for the system at frequency $\omega = \gamma$. Above it, the green curve shows the measured degree of squeezing through standard homodyne detection on the hybrid mechanical-optical modes at the same frequency. We see that, unlike the single-mode case, the two curves do not coincide at $\omega = \gamma$, the maximum achievable squeezing is inaccessible through standard HD at nonzero frequencies.

\subsection{\label{subsec:IVC} Two-mode detuned $\chi^{(3)}$-based OPO}
We now show an example where the asymmetry is rather in the term $G F$ as described by the second condition in Proposition \ref{prop GF}.
For that, we focus on the most generic two-mode detuned $\chi^{(3)}$-based OPO described by identical mode decay rates $\gamma$.
Below the oscillation threshold, the system can be described by a set of linear quantum Langevin equations in the form of Eq.~\eqref{eq:quantum_lang_r} where we have the $4\times4$ damping matrix $\Gamma = \gamma I_{4\times4}$ and the mode interaction matrix (Eq.~\ref{eq:M_cali}) with:
\begin{equation}
G = \begin{pmatrix}
g_{11} & g_{12} \\ 
g_{12}^{\ast} & g_{22}
\end{pmatrix}, \;
F = \begin{pmatrix}
0 & f_{12} \\ 
f_{12} & 0
\end{pmatrix}.
\label{eq:two_mode_opo_gf}
\end{equation}
This scenario occurs, for example, in microring resonators driven by one pump~\cite{Vaidya2020a}, or in the linearized dynamics driven slightly above threshold~\cite{Bensemhoun2024}. 

\begin{figure}[t!]
\centering
\includegraphics[width=1\columnwidth]{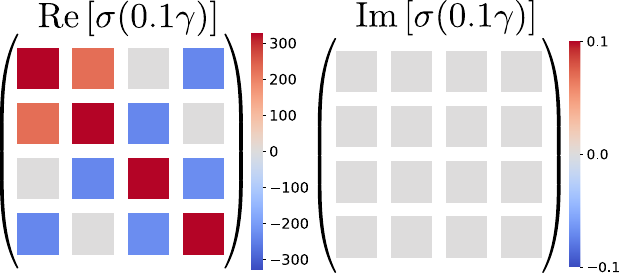}
\caption{Real spectral covariance matrix of a two-mode OPO at $\omega=0.1\gamma$ with mode-independent damping rates $\gamma$. We consider a parameter configuration reminiscent of broadband squeezing generation in microresonators in \cite{Vaidya2020a}. Mode-independent detunings are chosen such that $g_{11}= g_{22}=\gamma$, and without loss of generality, we consider $g_{12}=0$ and $f_{12}= 1.38i\gamma$. For these parameters, $(GF)^\text{T} = GF$. The figure displays the real (left) and imaginary (right) parts of the 16 matrix elements of the spectral covariance. The absence of imaginary components indicates a real spectral covariance.}
\label{fig:opo_two_covar_real}
\end{figure}

\begin{figure}[t!]
\centering
\includegraphics[width=1\columnwidth]{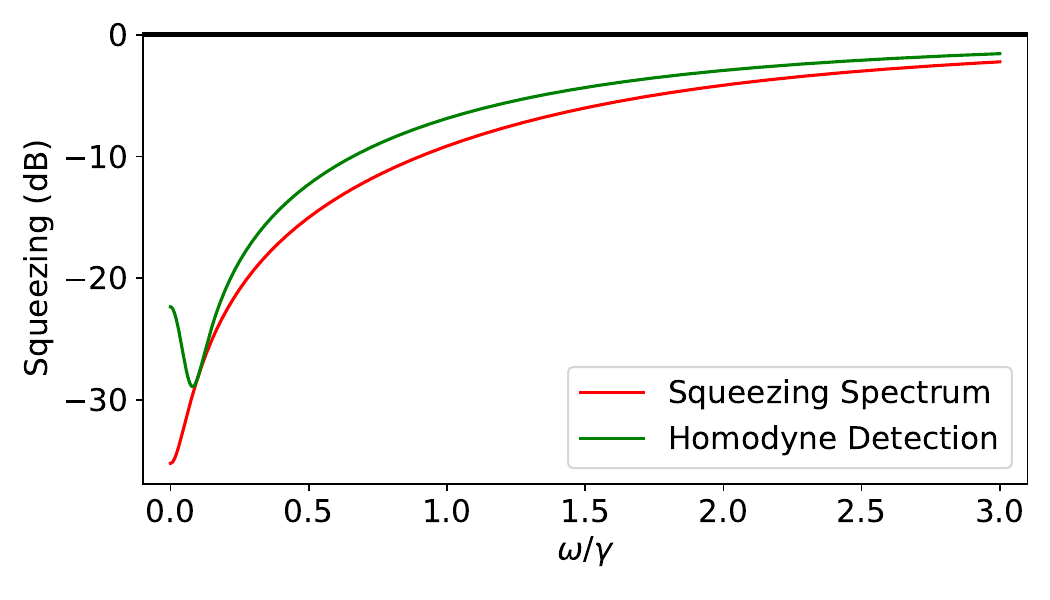}
\caption{Maximum achievable squeezing spectrum (red) for the two-mode OPO described by the complex spectral covariance matrix depicted in Fig.~\ref{fig:opo_two_covar_real} at $\omega = 0.1\gamma$. Spectrum measured via standard HD (green) at that frequency. Similar to the single-mode case, the two curves coincide at $\omega = 0.1\gamma$ , due to the real nature of the spectral covariance matrix. The maximum achievable squeezing is therefore accessible through standard HD at this frequency, and equivalently at all nonzero frequencies~\cite{Gouzien2023, Dioum2024}.}
\label{fig:opo_two_HD_real}
\end{figure}

When $g_{11} = g_{22}$, the product $GF$ is symmetric, resulting from Proposition ~\ref{prop GF} in a real spectral covariance matrix. This scenario corresponds, for example, to a parameter configuration reminiscent of broadband squeezing generation in microresonators~\cite{Vaidya2020a}. Notably, in their work, Vaidya et al. relied on HD to access the maximum optimal squeezing levels. For parameters, we choose mode-independent detunings such that $g_{11} = g_{22} = \gamma$, and without loss of generality, set $g_{12} = 0$ and $f_{12} = 1.38i\gamma$ to ensure that we are $5\%$ below the threshold. Fig.~\ref{fig:opo_two_covar_real} shows the real (left) and imaginary (right) parts of the 16 elements of the spectral covariance matrix of this configuration, at $\omega = 0.1\gamma$. The absence of imaginary components clearly indicates a real spectral covariance. Consequently, as for the single-mode case, we show in Fig.~\ref{fig:opo_two_HD_real}, the maximum achievable squeezing spectrum (red) and the spectrum measured via standard HD (green) coincide at $\omega = 0.1\gamma$. This result confirms our prediction and validates the approach taken by Vaidya et al., showing that they indeed could access the maximum optimal squeezing through HD in their experimental configuration.

%

\begin{figure}[t!]
\centering
\includegraphics[width=1\columnwidth]{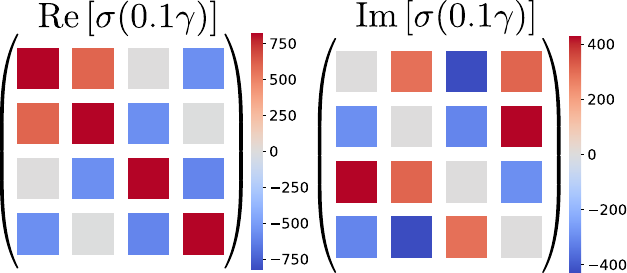}
\caption{Real (left) and imaginary (right) part of the 16 elements of the $4\times4$ complex spectral covariance matrix of  two-mode OPO at $\omega=0.1\gamma$ with mode independent damping rates $\gamma$. We considered also, for example, 
mode-dependent detuning $g_{11}=\gamma$ and $g_{22}=0.8\gamma$. We choose without loss of generality $g_{12}=0$ and $f_{12}= 1.31i\gamma$. For this choice of parameters $(GF)^\transp\neq GF$. We see the matrix exhibits nonzero imaginary components, 
indicating complex spectral covariance.}
\label{fig:opo_two_covar}
\end{figure}
\begin{figure}[t!]
\centering
\includegraphics[width=1\columnwidth]{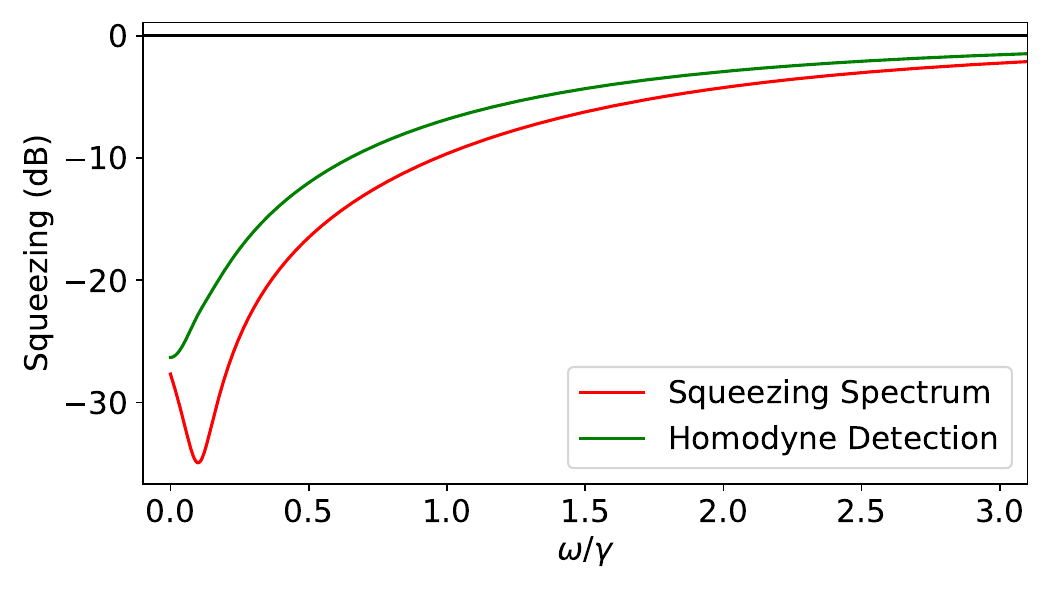}
\caption{Maximum achievable squeezing spectrum (red) for the two-mode OPO described by the complex spectral covariance matrix depicted in Fig.~\ref{fig:opo_two_covar} at $\omega = 0.1\gamma$. Spectrum measured via standard HD (green) at that frequency. Unlike the single-mode case, the two curves do not coincide at that frequency, due to the complex nature of the spectral covariance matrix. The maximum achievable squeezing is therefore inaccessible through standard HD at $\omega = 0.1\gamma$, and equivalently at all nonzero frequencies~\cite{Gouzien2023, Dioum2024}.}
\label{fig:opo_two_HD}
\end{figure}

However, slight asymmetries between modes can often be present such that $g_{11} \neq g_{22}$. This induces a breaking of the symmetry of the product $GF$ and could be generated in this type of system by, but not limited to, something as simple as dispersion-induced mode-dependent detuning. Therefore, despite equal damping rates ensures $\left[\Gamma,\mathcal{M}\right]=0$, the asymmetry introduced in the interaction matrices is sufficient to generate complex spectral covariance. 
We show in Fig.~\ref{fig:opo_two_covar}, for a chosen nonzero frequency ($\omega=0.1\gamma$), the real (on the left) and 
imaginary (on the right) parts of the 16 elements contained in the $4\times4$ spectral covariance $\sigma_{\mathrm{out}}(\omega)$ 
for a given set of parameters of the OPO model, $5\%$ below threshold ($g_{11} = \gamma$,  $g_{22}=0.8 \gamma$,  $g_{12}=0$,  $f_{12}=1.31i\gamma$).

The consequence is that optimal levels of squeezing cannot be fully accessed through standard HD. This is illustrated with Fig.~\ref{fig:opo_two_HD}: the degree of squeezing as measured by HD (solid-green curve) doesn't reach the maximum achievable squeezing (solid-red curve) at the chosen frequency $\omega = 0.1\gamma$. This also remains true for every other nonzero frequency.

\subsection{Single-mode from reduced three-mode system}


\begin{figure}
    \centering
    \includegraphics[width=\linewidth]{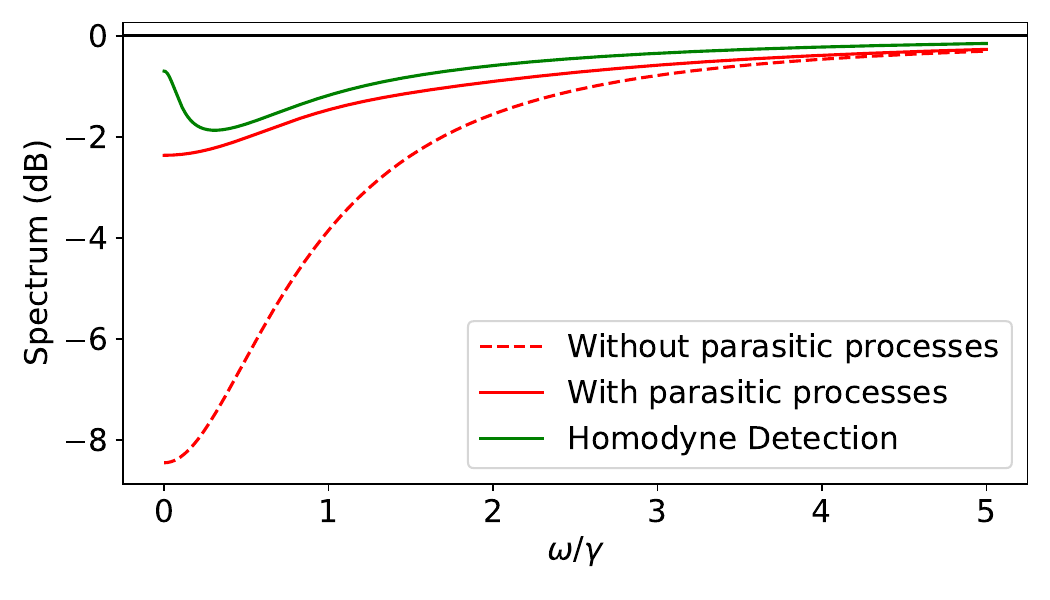}
    \caption{Squeezing spectrum of the signal mode ($s$) with (solid red) and without (dashed red) parasitic processes compared to the squeezing obtained through standard HD (green) optimized at a chosen frequency $\omega = 0.5 \gamma$. We considered  a dual pumped microresonator system for broadband squeezing generation~\cite{Seifoory2022} where $G$ and $F$ are characterized by Eq.~\eqref{eq_G_F_Seifoory} with $\beta_{p1}\beta_{p2} = 0.46 \gamma$ and $|\beta_{p1}|^2 + |\beta_{p2}|^2 - \Delta = 0.15\gamma$. We see that a large portion of the signal squeezing is reduced due to the presence of parasitic processes. However, the squeezing measured via HD (green) still does not match at the optimized frequency, the maximum reamining squeezing. In particular, at this frequency, $13\%$ of the squeezing remaining after reduction from the parasitic processes is inaccessible to HD schemes.}
    \label{fig:seifoory}
\end{figure}

In the work of Seifoory et al. \cite{Seifoory2022}, the authors analyze the generation of squeezed light through SFWM in a dual-pumped microring resonator, focusing on the detection of squeezing in the signal mode $s$. Their work highlights how, due to unwanted nonlinear optical processes, squeezing is reduced. Our analysis predicts, however, that even when accounting for such parasitic processes, there exists additional hidden and detectable squeezing that remains inaccessible to conventional HD techniques.

Consider their three-mode system $m$, $s$ and $n$ with two strong classical pumps, the Hamiltonian for the nonlinear processes includes interactions 
such self-phase modulation (SPM), cross-phase modulation (XPM), dual-pump spontaneous four-wave mixing (DP-SFWM), single-pump spontaneous four-wave mixing (SP-SFWM), Bragg-scattering four-wave mixing (BS-FWM), and hyperparametric spontaneous four-wave mixing (HP-SFWM), each contributing to the overall dynamics of the system. The main process for squeezing generation is DP-SFWM and the main process responsable squeezing reduction is SP-SFWM, which injects uncorrelated photons into the signal resonance. BS-FWM and HP-SFWM also contribute to squeezing degradation, but to a lesser extent than SP-SFWM.

In the classical pumps approximation and moving to the interaction picture, the dynamics of the system can be described the quantum Langevin equations in Eq.~\ref{eq:quantum_lang_r} where $\Gamma = \gamma I_{6\times6}$ and the matrices $G$ and $F$ are :

\begin{widetext}
\begin{align}
G &= - \begin{pmatrix}
|\beta_{p1}|^2 + |\beta_{p2}|^2 - \Delta & \beta_{p1}\beta_{p2}^{\ast} & 0 \\
\beta_{p1}^\ast \beta_{p2} & |\beta_{p1}|^2 + |\beta_{p2}|^2 - \Delta & \beta_{p1}\beta_{p2}^\ast \\
0 & \beta_{p1}^{\ast}\beta_{p2} & |\beta_{p1}|^2 + |\beta_{p2}|^2 - \Delta
\end{pmatrix}, 
\nonumber\\
F &= - \begin{pmatrix}
0 & \beta_{p1}^2 & 2\beta_{p1}\beta_{p2} \\
\beta_{p1}^2 & \beta_{p1}\beta_{p2} & \beta_{p2}^2 \\
2\beta_{p1}\beta_{p2} & \beta_{p2}^2 & 0
\end{pmatrix}.
\label{eq_G_F_Seifoory}
\end{align}
\end{widetext}

Here, $\beta_{p1}$ and $\beta_{p2}$ are proportional to the complex classical pump amplitudes $\beta'_{j}$ with $\beta_{j} = \sqrt{2\Lambda}\beta'_{j}$, $j$ in $ \{p_1, p_2\}$ and $\Lambda$ represents the nonlinear coefficient quantifying the strength of the nonlinear optical processes. We have considered the same detuning $\Delta$ to the microring resonances for both pumps.

To verify Proposition \ref{prop GF}, we observe that the product $GF$ is not symmetric, leading a complex spectral covariance matrix for the three-mode system. This complexity persists even when we trace down the system to single mode, focusing only on the signal mode. Correspondingly, as shown in previous examples and demonstrated in \cite{Dioum2024}, even after the reduction of squeezing due to parasitic processes, HD still cannot fully access all the remaining squeezing still available in the $s$ mode. This is illustrated in Fig.~\ref{fig:seifoory} where we show the squeezing spectrum of the signal mode ($s$) under various conditions for the dual-pumped microresonator system described by Seifoory et al.~\cite{Seifoory2022} with chosen parameters $\beta_{p1}\beta_{p2} = 0.46 \gamma$ and $|\beta_{p1}|^2 + |\beta_{p2}|^2 - \Delta = 0.15\gamma$. The solid (dashed) red line represents the spectrum without (with) parasitic processes and the green line indicates the squeezing obtained through standard HD, optimized at $\omega = 0.5\gamma$. As shown by the difference between the solid and dashed red lines, it's evident that a significant portion of the signal squeezing is reduced when parasitic processes are present. Furthermore, the squeezing measured via HD (green line) accounts only for $83\%$ of the maximum available remaining squeezing, in the presence of parasitic processes at the optimized frequency $\omega = 0.5\gamma$. Suboptimal squeezing measurements are therefore not solely due to parasitic processes, as suggested by Seifoory et al. Rather, our criteria predict $13\%$ of the detectable squeezing is hidden from HD and can be revealed by the use of novel detection schemes~\cite{Barbosa2013a, Buchmann2016, Dioum2024}.

The squeezing spectrum without parasitic processes shown in dashed-red in Fig.~\ref{fig:seifoory} was obtained using the strategy described by Zhang et al. \cite{Zhang2021}. Their approach employs a photonic molecule consisting of two coupled microring resonators on an integrated nanophotonic chip. This design selectively hybridizes only the modes involved in unwanted processes, effectively suppressing parasitic processes such as DP-SFWM, BS-FWM, and HP-SFWM.
In particular, our criteria show that this method also addresses the issue of hidden squeezing. This can be demonstrated by examining the new interaction matrices of the new system when parasitic processes are suppressed: $G = (|\beta_{p1}|^2 + |\beta_{p2}|^2 - \Delta)I_{3\times3}$ and $F = -\mathrm{diag}\{0, \beta_{p1}\beta_{p2}, 0\}$. The symmetry of the product $GF$ implies that homodyne detection can access the optimal squeezing in cases where these parasitic processes are suppressed.

\section{\label{sec:V} Conclusion}

The nature of the spectral covariance matrix characterizing the output quantum state of a cavity-based optical system plays a crucial role in the choice of the measurement scheme. When the spectral covariance matrix is real, standard HD is sufficient for complete characterization. However, in the case where it's complex, HD becomes insufficient, and other detection methods such as ``resonator detection"~\cite{Barbosa2013b,Barbosa2013b}, ``synodine detection~"\cite{Buchmann2016} or ``interferometers with memory effect"~\cite{Dioum2024} become necessary to fully capture all correlations. Besides, an ability to control the nature of spectral covariances based on the system's parameters, can allow for the engineering of these systems to tailor or avoid this complex nature.

In this work, we have derived key conditions under which the output spectral covariance matrix $\sigma_{\mathrm{out}}(\omega)$ of a cavity-based optical quantum system, characterized by the most general quadratic Hamiltonian, is constrained to be either real- or complex-valued.
Specifically, we showed that beyond the trivial cases $F=0$ and $\mathcal{M}$ diagonal, the spectral covariance matrix is real for systems with mode-independent damping rates and when the product of the interaction matrices $GF$ is symmetric. Failing to satisfy either of these conditions necessarily results in a complex spectral covariance matrix, indicating the presence of hidden correlations.
This connects the structure of the fundamental mode interaction matrices $F$ and $G$ to the real or complex nature of the spectral covariance matrix. 

We showed examples that elucidate the transition from real spectral covariance matrices accessible via standard techniques to complex spectral covariance where hidden correlations are present.
We gained insights into types of interactions that prohibit or permit complex spectral behavior, demarcating the boundary between standard and hidden squeezing.

\begin{acknowledgments}
 We are grateful for discussions to Carlos Navarrete-Benlloch about the spectral covariance matrix and to Alessandro Pugliese about analytic decompositions.
 
We acknowledge funding from the Plan France 2030 through the project ANR-22-PETQ-0013. Virginia D'Auria acknowledges financial support from the Institut Universitaire de France (IUF) and the project SPHIFA (ANR-20-CE47-0012). Giuseppe Patera acknowledges financial support from the Agence Nationale de la Recherche through LABEX CEMPI (ANR-11-LABX-0007) and I-SITE ULNE (ANR-16-IDEX-0004).
\end{acknowledgments}

\appendix

\section{\label{ap:I} Complex spectral covariance matrix}

Here we details more lenghtier derivations of properties in Section \ref{sec:IIA}.
The covariance matrix for Gaussian states where we have considered zeros mean value can be defined as Eq.~\eqref{eq:covar_temp} in the main text.
Using the Fourier transform transfer of the covariance matrix elements-wise as in Eq.~\eqref{eq:covar_spect} and pluging the definition of the 
inverse Fourier transform of $\sigma_{m,n}(\tau)$, one can show the form of the elements of the spectral covariance matrix:

\begin{align}
\sigma_{m,n}(\omega)&=\int d\tau \mathrm{e}^{i \omega \tau}
\left[\int \frac{d\omega'}{\sqrt{2\pi}} \mathrm{e}^{i \omega' \tau} 
\sigma_{m,n}(\omega,\omega')
\right]
\nonumber\\
&= \int d\omega' 
\left[\int \frac{d\tau}{\sqrt{2\pi}}\mathrm{e}^{i (\omega + \omega') \tau} 
\right ]
\sigma_{m,n}(\omega,\omega')
\nonumber\\
& = \int d\omega' \delta(\omega + \omega')
\sigma_{m,n}(\omega,\omega')
\nonumber\\
\sigma_{m,n}(\omega)& = \sigma_{m,n}(\omega,-\omega),
\label{seq:covar_spect_2_mn} 
\end{align}

yielding the spectral covariance matrix in Eq. (\ref{eq:covar_spect}).

The spectral quadrature operators in $\hat{\mathbf{R}}(\omega)$ are defined as the Fourier transforms of the temporal counterparts (\ref{eq:fourier_R}). One can easily show from this definition that :

\begin{align}
[\hat{\mathbf{R}}(\omega)]^{\dagger} 
&= \left[
\int \frac{dt}{\sqrt{2\pi}} \mathrm{e}^{-i \omega t} \hat{\mathbf{R}}(t)
\right]^{\dagger}
\nonumber\\
&= 
\int \frac{dt}{\sqrt{2\pi}} \mathrm{e}^{i \omega t} \hat{\mathbf{R}}^{\dagger}(t)
\nonumber\\
&=
\int \frac{dt}{\sqrt{2\pi}} \mathrm{e}^{i \omega t} \hat{\mathbf{R}}(t)
\nonumber\\
[\hat{\mathbf{R}}(\omega)]^{\dagger} &=
\hat{\mathbf{R}}(-\omega),
\label{seq:spect_quadr_compl}
\end{align}

which gives the following spectral commutation relations

\begin{align}
[\hat{\mathbf{R}} (\omega),\hat{\mathbf{R}}^{\dagger} (\omega')]
&=\int \frac{dt_1}{\sqrt{2\pi}} \mathrm{e}^{-i \omega t_1}
\nonumber \\
&\int \frac{dt_2}{\sqrt{2\pi}} \mathrm{e}^{-i \omega' t_2}
[\hat{\mathbf{R}} (t_1),\hat{\mathbf{R}}^{\dagger} (t_2)]
\nonumber\\
&= i\Omega\int \frac{dt_1}{\sqrt{2\pi}}
\nonumber \\
&\int \frac{dt_2}{\sqrt{2\pi}}
\mathrm{e}^{-i (\omega t_1 + \omega' t_2)} \delta(t_1-t_2)
\nonumber\\
&=i\Omega\int \frac{dt_1}{\sqrt{2\pi}} \mathrm{e}^{-i (\omega + \omega') t_1},
\nonumber\\
[\hat{\mathbf{R}} (\omega),\hat{\mathbf{R}}^{\dagger} (\omega')]
& = i\Omega\delta(\omega+\omega').
\label{seq:commut_spect}
\end{align}

\section{\label{ap:II} Complex transfer function}

We can explicitly write out the real and imaginary parts of $S(\omega) = S_{\mathrm{R}}(\omega) + iS_{\mathrm{I}}(\omega)$,
Where $S_{\mathrm{R}}(\omega)) = (S(\omega) + S(-\omega))/2$ is the symmetric part and $S_{\mathrm{I}}(\omega) = (S(\omega) - S(-\omega))/2i$ is the antisymmetric part.

More explicitly, 
\begin{align}
S_{\mathrm{R}}(\omega)
=\frac{1}{2} \Big( S(\omega&) + S^{\ast}(\omega)\Big) \nonumber\\
=\frac{1}{2} \Big( S(\omega&) + S(-\omega)\Big) \nonumber\\
=\frac{\sqrt{2\Gamma}}{2}\Big[
(&i\omega\mathbb{I}+\Gamma-\mathcal{M})^{-1} + \nonumber \\
(&-i\omega\mathbb{I}+\Gamma-\mathcal{M})^{-1}
\Big]\sqrt{2\Gamma}-\mathbb{I} \nonumber\\
=\frac{\sqrt{2\Gamma}}{2}\Big[
(&-i\omega\mathbb{I}+\Gamma-\mathcal{M})
 \mathrm{Inv}(\omega)+ \nonumber\\
(&i\omega\mathbb{I}+\Gamma-\mathcal{M})
\mathrm{Inv}(\omega)
\Big]\sqrt{2\Gamma}-\mathbb{I}  \nonumber\\
S_{\mathrm{R}}(\omega)
=\sqrt{2\Gamma}\Big[
(&\Gamma-\mathcal{M})
\mathrm{Inv}(\omega)
\Big]\sqrt{2\Gamma}-\mathbb{I}
\label{seq:s_r}
\end{align}

where we have considered that the product:
\begin{align}
\mathrm{Inv} (\omega)
 &= (-i\omega\mathbb{I}+\Gamma-\mathcal{M})^{-1}
(i\omega\mathbb{I}+\Gamma-\mathcal{M})^{-1} 
\nonumber\\&=
(i\omega\mathbb{I}+\Gamma-\mathcal{M})^{-1}
(-i\omega\mathbb{I}+\Gamma-\mathcal{M})^{-1}
 \nonumber\\
&=
\Big((i\omega\mathbb{I}+\Gamma-\mathcal{M})
(-i\omega\mathbb{I}+\Gamma-\mathcal{M})\Big)^{-1}
 \nonumber\\
\mathrm{Inv}(\omega)&=
\Big( \omega^{2}\mathbb{I} +(\Gamma-\mathcal{M})^{2}\Big)^{-1}.
\end{align}

We can further simplify Eq.~\eqref{seq:s_r} by absorbing the $\mathbb{I}$ in the expresion:

\begin{align}
S_{\mathrm{R}}(\omega)=& 
\sqrt{2\Gamma}\Big[
(\Gamma-\mathcal{M})
\mathrm{Inv}(\omega)
\Big]\sqrt{2\Gamma}  \nonumber\\ 
&- \frac{1}{\sqrt{2\Gamma}} 
\mathrm{Inv}^{-1} (\omega) \mathrm{Inv}(\omega)
\sqrt{2\Gamma}
 \nonumber\\ 
=& \frac{1}{\sqrt{2\Gamma}}
\Big[
2\Gamma(\Gamma-\mathcal{M})- \mathrm{Inv}^{-1} (\omega)
\Big]
\mathrm{Inv}(\omega) \sqrt{2\Gamma}
 \nonumber\\ 
=& \frac{1}{\sqrt{2\Gamma}}
\Big[
2\Gamma(\Gamma-\mathcal{M})- 
[\omega^{2}\mathbb{I} +(\Gamma-\mathcal{M})^{2}]
\Big]
\nonumber\\ 
&\;\mathrm{Inv}(\omega) \sqrt{2\Gamma}
 \nonumber\\ 
=& \frac{1}{\sqrt{2\Gamma}}
\Big[
(\Gamma+\mathcal{M})(\Gamma-\mathcal{M})
-\omega^{2}\mathbb{I}
\Big]
\mathrm{Inv}(\omega) \sqrt{2\Gamma}
 \nonumber\\ 
S_{\mathrm{R}}(\omega)=&\frac{1}{\sqrt{2\Gamma}}
\Big(\Gamma^2+[\mathcal{M}, \Gamma]-\mathcal{M}^2 -\omega^2 \mathbb{I}\Big)
\mathrm{Inv}(\omega)
\sqrt{2\Gamma}.
\label{seq:s_r_2}
\end{align}

And similarly

\begin{align}
S_{\mathrm{I}}(\omega)
&=\frac{1}{2i}\left(S(\omega) + S^{\ast}(\omega)\right)
\nonumber\\
&=\frac{1}{2i}\left(S(\omega) + S(-\omega)\right)
\nonumber\\
&=\frac{\sqrt{2\Gamma}}{i}\Big[
-i\omega
\mathrm{Inv}(\omega)
\Big]\sqrt{2\Gamma}
\nonumber\\
S_{\mathrm{I}}(\omega)&=\sqrt{2\Gamma}\Big[
-\omega
\mathrm{Inv}(\omega)
\Big]\sqrt{2\Gamma}.
\label{seq:s_i}
\end{align}

The product $S_{\mathrm{R}}(\omega)S_{\mathrm{I}}^{\transp}(\omega)$ writes:

\begin{align}
S_{\mathrm{R}}(\omega)S_{\mathrm{I}}^{\transp}(\omega)= 
& \frac{-2\omega}{\sqrt{2\Gamma}} 
\Big(\Gamma^2 + [\mathcal{M}, \Gamma]-\mathcal{M}^2 -\omega^2 \mathbb{I}\Big)
\nonumber \\
&\Big[\mathrm{Inv}(\omega)
\Gamma
\mathrm{Inv}^{\transp}(\omega)
\Big]\sqrt{2\Gamma}
\nonumber\\
S_{\mathrm{R}}(\omega)S_{\mathrm{I}}^{\transp}(\omega)= 
&\frac{-2\omega}{\sqrt{2\Gamma}}
A(\omega)B(\omega)
\sqrt{2\Gamma},
\label{seq:product}
\end{align}

where $A(\omega)=\Gamma^2+[\mathcal{M}, \Gamma]-\mathcal{M}^2-\omega^2 \mathbb{I}$ and $B(\omega) = \mathrm{Inv}(\omega) \Gamma \mathrm{Inv}^{\transp}(\omega)$. Note that $B(\omega)$ is a symmetric matrix $B^{\transp}(\omega) = B(\omega)$. One can also write the transpose of this Eq.~\ref{seq:product}, $S_{\mathrm{I}}(\omega)S_{\mathrm{R}}^{\transp}(\omega)$ which writes:

\begin{align}
S_{\mathrm{I}}(\omega)S_{\mathrm{R}}^{\transp}(\omega)&= 
-2\omega \sqrt{2\Gamma}
B(\omega)A(\omega)^{\transp}
\frac{1}{\sqrt{2\Gamma}}.
\label{seq:transpose_product}
\end{align}

To find out when the product is $S_{\mathrm{R}}(\omega)S_{\mathrm{I}}^{\transp}(\omega)$ is symmetric, we write the difference between Eq.~\eqref{seq:product} and Eq.~\eqref{seq:transpose_product}:

\begin{align}
S_{\mathrm{R}}(\omega)S_{\mathrm{I}}^{\transp}(\omega)-S_{\mathrm{I}}(\omega)S_{\mathrm{R}}^{\transp}(\omega) = 
&\frac{-2\omega}{\sqrt{2\Gamma}}
A(\omega)B(\omega)
\frac{2\Gamma}{\sqrt{2\Gamma}}
\nonumber\\
+&2\omega \frac{2\Gamma}{\sqrt{2\Gamma}}
B(\omega)A^{\transp}(\omega)
\frac{1}{\sqrt{2\Gamma}},
\nonumber\\
S_{\mathrm{R}}(\omega)S_{\mathrm{I}}^{\transp}(\omega)-S_{\mathrm{I}}(\omega)S_{\mathrm{R}}^{\transp}(\omega) = 
&\frac{-4\omega}{\sqrt{2\Gamma}}
\Big(A(\omega)B(\omega)\Gamma 
\nonumber\\
- &\Gamma B(\omega) A^{\transp}(\omega)\Big)
\frac{1}{\sqrt{2\Gamma}}.
\label{seq:difference}
\end{align}
\section{\label{ap:III} Pair-wise commutation between $\Gamma$, $A(\omega)$ and $B(\omega)$}
In the section, we want to show that for $\Gamma$, $A(\omega)$ and $B(\omega)$ to commute pair-wise, one must ensure $[\Gamma, \mathcal{M}]=0$.

First, for $\Gamma$ to commute with $A(\omega)$, it must commute with $\Gamma \mathcal{M}$, $\mathcal{M}\Gamma $ and $\mathcal{M}^2$. One can easily show that $[\Gamma, \mathcal{M}\Gamma]=0$ and $[\Gamma, \Gamma\mathcal{M}]=0$ are both equivalent to $[\Gamma, \mathcal{M}]=0$, which also ensures that $[\Gamma, \mathcal{M}^2]=0$. Therefore for $\Gamma$ to commute with $A(\omega)$, it must commute with $\mathcal{M}$.

It is also easy to show that since $\Gamma$ commutes with $\mathcal{M}$ and therefore with $\mathcal{M}^2$ and $\mathcal{M}\Gamma$, it consequently also commutes with $(\Gamma-\mathcal{M})^2$ and  therefore with both $\mathrm{Inv}(\omega)$ and $\mathrm{Inv}^{\transp}(\omega)$. This ensures that for $\Gamma$ to commute with both $A(\omega)$ and $B(\omega)$, it must commute with $\mathcal{M}$.

To ensure pair-wise commutation between $\Gamma$, $A(\omega)$ and $B(\omega)$, it is left to show when $[A(\omega),B(\omega)] = 0$. We can also notice that when $\Gamma$ commutes with $\mathcal{M}$, all the previous commutations we mentioned early hold true also for $\Gamma^2$ and $\mathcal{M}^2$. In particular, they both commute with 
$\mathrm{Inv}(\omega)$ and therefore with $\mathrm{Inv}^{\transp}(\omega)$. 

We can conclude that for $\Gamma$, $A(\omega)$ and $B(\omega)$ commute pair-wise when $[\Gamma, \mathcal{M}]=0$.

\section{\label{ap:IV} Symmetry of the interaction matrices}

To gain further insights from  $M^2$ and using the condition of hermicity of G and symmetry of F, we can write it explicitly :

\begin{align}
M^{2} &= \begin{pmatrix}
G^{2}-FF^{\ast} & GF-FG^{\ast} \\
G^{\ast}F^{\ast}-F^{\ast}G & G^{\ast2}-F^{\ast}F
\end{pmatrix}
 \nonumber \\
&= \begin{pmatrix}
G^{2}-FF^{\ast} & GF-FG^{\ast} \\
-(GF-FG^{\ast})^{\dagger} & (G^{2}-FF^{\ast})^{\transp}
\end{pmatrix}.
\label{seq:M^2}
\end{align}


\bibliography{library}

\begin{thebibliography}{56}%
\makeatletter
\providecommand \@ifxundefined [1]{%
 \@ifx{#1\undefined}
}%
\providecommand \@ifnum [1]{%
 \ifnum #1\expandafter \@firstoftwo
 \else \expandafter \@secondoftwo
 \fi
}%
\providecommand \@ifx [1]{%
 \ifx #1\expandafter \@firstoftwo
 \else \expandafter \@secondoftwo
 \fi
}%
\providecommand \natexlab [1]{#1}%
\providecommand \enquote  [1]{``#1''}%
\providecommand \bibnamefont  [1]{#1}%
\providecommand \bibfnamefont [1]{#1}%
\providecommand \citenamefont [1]{#1}%
\providecommand \href@noop [0]{\@secondoftwo}%
\providecommand \href [0]{\begingroup \@sanitize@url \@href}%
\providecommand \@href[1]{\@@startlink{#1}\@@href}%
\providecommand \@@href[1]{\endgroup#1\@@endlink}%
\providecommand \@sanitize@url [0]{\catcode `\\12\catcode `\$12\catcode `\&12\catcode `\#12\catcode `\^12\catcode `\_12\catcode `\%12\relax}%
\providecommand \@@startlink[1]{}%
\providecommand \@@endlink[0]{}%
\providecommand \url  [0]{\begingroup\@sanitize@url \@url }%
\providecommand \@url [1]{\endgroup\@href {#1}{\urlprefix }}%
\providecommand \urlprefix  [0]{URL }%
\providecommand \Eprint [0]{\href }%
\providecommand \doibase [0]{https://doi.org/}%
\providecommand \selectlanguage [0]{\@gobble}%
\providecommand \bibinfo  [0]{\@secondoftwo}%
\providecommand \bibfield  [0]{\@secondoftwo}%
\providecommand \translation [1]{[#1]}%
\providecommand \BibitemOpen [0]{}%
\providecommand \bibitemStop [0]{}%
\providecommand \bibitemNoStop [0]{.\EOS\space}%
\providecommand \EOS [0]{\spacefactor3000\relax}%
\providecommand \BibitemShut  [1]{\csname bibitem#1\endcsname}%
\let\auto@bib@innerbib\@empty
\bibitem [{\citenamefont {Braunstein}\ and\ \citenamefont {{Van Loock}}(2005)}]{Braunstein2005a}%
  \BibitemOpen
  \bibfield  {author} {\bibinfo {author} {\bibfnamefont {L.~S.}\ \bibnamefont {Braunstein}}\ and\ \bibinfo {author} {\bibfnamefont {P.}~\bibnamefont {{Van Loock}}},\ }\bibfield  {title} {\bibinfo {title} {{Quantum information with continuous variables}},\ }\href {https://doi.org/10.1103/RevModPhys.77.513} {\bibfield  {journal} {\bibinfo  {journal} {Reviews of Modern Physics}\ }\textbf {\bibinfo {volume} {77}},\ \bibinfo {pages} {513} (\bibinfo {year} {2005})},\ \Eprint {https://arxiv.org/abs/0410100} {arXiv:0410100 [quant-ph]} \BibitemShut {NoStop}%
\bibitem [{\citenamefont {Adesso}\ \emph {et~al.}(2014)\citenamefont {Adesso}, \citenamefont {Ragy},\ and\ \citenamefont {Lee}}]{Adesso2014a}%
  \BibitemOpen
  \bibfield  {author} {\bibinfo {author} {\bibfnamefont {G.}~\bibnamefont {Adesso}}, \bibinfo {author} {\bibfnamefont {S.}~\bibnamefont {Ragy}},\ and\ \bibinfo {author} {\bibfnamefont {A.~R.}\ \bibnamefont {Lee}},\ }\bibfield  {title} {\bibinfo {title} {{Continuous variable quantum information: Gaussian states and beyond}},\ }\bibfield  {journal} {\bibinfo  {journal} {Open Systems and Information Dynamics}\ }\textbf {\bibinfo {volume} {21}},\ \href {https://doi.org/10.1142/S1230161214400010} {10.1142/S1230161214400010} (\bibinfo {year} {2014}),\ \Eprint {https://arxiv.org/abs/1401.4679} {arXiv:1401.4679} \BibitemShut {NoStop}%
\bibitem [{\citenamefont {Menicucci}\ \emph {et~al.}(2006)\citenamefont {Menicucci}, \citenamefont {{Van Loock}}, \citenamefont {Gu}, \citenamefont {Weedbrook}, \citenamefont {Ralph},\ and\ \citenamefont {Nielsen}}]{Menicucci2006}%
  \BibitemOpen
  \bibfield  {author} {\bibinfo {author} {\bibfnamefont {N.~C.}\ \bibnamefont {Menicucci}}, \bibinfo {author} {\bibfnamefont {P.}~\bibnamefont {{Van Loock}}}, \bibinfo {author} {\bibfnamefont {M.}~\bibnamefont {Gu}}, \bibinfo {author} {\bibfnamefont {C.}~\bibnamefont {Weedbrook}}, \bibinfo {author} {\bibfnamefont {T.~C.}\ \bibnamefont {Ralph}},\ and\ \bibinfo {author} {\bibfnamefont {M.~A.}\ \bibnamefont {Nielsen}},\ }\bibfield  {title} {\bibinfo {title} {{Universal quantum computation with continuous-variable cluster states}},\ }\href {https://doi.org/10.1103/PhysRevLett.97.110501} {\bibfield  {journal} {\bibinfo  {journal} {Physical Review Letters}\ }\textbf {\bibinfo {volume} {97}},\ \bibinfo {pages} {13} (\bibinfo {year} {2006})},\ \Eprint {https://arxiv.org/abs/0605198} {arXiv:0605198 [quant-ph]} \BibitemShut {NoStop}%
\bibitem [{\citenamefont {Gu}\ \emph {et~al.}(2009)\citenamefont {Gu}, \citenamefont {Weedbrook}, \citenamefont {Menicucci}, \citenamefont {Ralph},\ and\ \citenamefont {{Van Loock}}}]{Gu2009}%
  \BibitemOpen
  \bibfield  {author} {\bibinfo {author} {\bibfnamefont {M.}~\bibnamefont {Gu}}, \bibinfo {author} {\bibfnamefont {C.}~\bibnamefont {Weedbrook}}, \bibinfo {author} {\bibfnamefont {N.~C.}\ \bibnamefont {Menicucci}}, \bibinfo {author} {\bibfnamefont {T.~C.}\ \bibnamefont {Ralph}},\ and\ \bibinfo {author} {\bibfnamefont {P.}~\bibnamefont {{Van Loock}}},\ }\bibfield  {title} {\bibinfo {title} {{Quantum computing with continuous-variable clusters}},\ }\href {https://doi.org/10.1103/PhysRevA.79.062318} {\bibfield  {journal} {\bibinfo  {journal} {Physical Review A - Atomic, Molecular, and Optical Physics}\ }\textbf {\bibinfo {volume} {79}},\ \bibinfo {pages} {1} (\bibinfo {year} {2009})},\ \Eprint {https://arxiv.org/abs/0903.3233} {arXiv:0903.3233} \BibitemShut {NoStop}%
\bibitem [{\citenamefont {Giovannetti}\ \emph {et~al.}(2004)\citenamefont {Giovannetti}, \citenamefont {Lloyd},\ and\ \citenamefont {Maccone}}]{Giovannetti2004}%
  \BibitemOpen
  \bibfield  {author} {\bibinfo {author} {\bibfnamefont {V.}~\bibnamefont {Giovannetti}}, \bibinfo {author} {\bibfnamefont {S.}~\bibnamefont {Lloyd}},\ and\ \bibinfo {author} {\bibfnamefont {L.}~\bibnamefont {Maccone}},\ }\bibfield  {title} {\bibinfo {title} {Quantum-enhanced measurements: Beating the standard quantum limit},\ }\href {https://doi.org/10.1126/science.1104149} {\bibfield  {journal} {\bibinfo  {journal} {Science}\ }\textbf {\bibinfo {volume} {306}},\ \bibinfo {pages} {1330} (\bibinfo {year} {2004})},\ \Eprint {https://arxiv.org/abs/https://www.science.org/doi/pdf/10.1126/science.1104149} {https://www.science.org/doi/pdf/10.1126/science.1104149} \BibitemShut {NoStop}%
\bibitem [{\citenamefont {Ferraro}\ \emph {et~al.}(2005)\citenamefont {Ferraro}, \citenamefont {Olivares},\ and\ \citenamefont {Paris}}]{Ferraro2005}%
  \BibitemOpen
  \bibfield  {author} {\bibinfo {author} {\bibfnamefont {A.}~\bibnamefont {Ferraro}}, \bibinfo {author} {\bibfnamefont {S.}~\bibnamefont {Olivares}},\ and\ \bibinfo {author} {\bibfnamefont {M.~G.~A.}\ \bibnamefont {Paris}},\ }\href {http://arxiv.org/abs/quant-ph/0503237} {\emph {\bibinfo {title} {{Gaussian states in continuous variable quantum information}}}}\ (\bibinfo {year} {2005})\ \Eprint {https://arxiv.org/abs/0503237} {arXiv:0503237 [quant-ph]} \BibitemShut {NoStop}%
\bibitem [{\citenamefont {Weedbrook}\ \emph {et~al.}(2012)\citenamefont {Weedbrook}, \citenamefont {Pirandola}, \citenamefont {Garc{\'{i}}a-Patr{\'{o}}n}, \citenamefont {Cerf}, \citenamefont {Ralph}, \citenamefont {Shapiro},\ and\ \citenamefont {Lloyd}}]{Weedbrook2012}%
  \BibitemOpen
  \bibfield  {author} {\bibinfo {author} {\bibfnamefont {C.}~\bibnamefont {Weedbrook}}, \bibinfo {author} {\bibfnamefont {S.}~\bibnamefont {Pirandola}}, \bibinfo {author} {\bibfnamefont {R.}~\bibnamefont {Garc{\'{i}}a-Patr{\'{o}}n}}, \bibinfo {author} {\bibfnamefont {N.~J.}\ \bibnamefont {Cerf}}, \bibinfo {author} {\bibfnamefont {T.~C.}\ \bibnamefont {Ralph}}, \bibinfo {author} {\bibfnamefont {J.~H.}\ \bibnamefont {Shapiro}},\ and\ \bibinfo {author} {\bibfnamefont {S.}~\bibnamefont {Lloyd}},\ }\bibfield  {title} {\bibinfo {title} {{Gaussian quantum information}},\ }\href {https://doi.org/10.1103/RevModPhys.84.621} {\bibfield  {journal} {\bibinfo  {journal} {Reviews of Modern Physics}\ }\textbf {\bibinfo {volume} {84}},\ \bibinfo {pages} {621} (\bibinfo {year} {2012})},\ \Eprint {https://arxiv.org/abs/1110.3234} {arXiv:1110.3234} \BibitemShut {NoStop}%
\bibitem [{\citenamefont {Braunstein}\ and\ \citenamefont {Kimble}(1998)}]{Braunstein1998a}%
  \BibitemOpen
  \bibfield  {author} {\bibinfo {author} {\bibfnamefont {S.~L.}\ \bibnamefont {Braunstein}}\ and\ \bibinfo {author} {\bibfnamefont {H.~J.}\ \bibnamefont {Kimble}},\ }\bibfield  {title} {\bibinfo {title} {Teleportation of continuous quantum variables},\ }\href {https://doi.org/10.1103/PhysRevLett.80.869} {\bibfield  {journal} {\bibinfo  {journal} {Phys. Rev. Lett.}\ }\textbf {\bibinfo {volume} {80}},\ \bibinfo {pages} {869} (\bibinfo {year} {1998})}\BibitemShut {NoStop}%
\bibitem [{\citenamefont {Furusawa}\ \emph {et~al.}(1998{\natexlab{a}})\citenamefont {Furusawa}, \citenamefont {Sørensen}, \citenamefont {Braunstein}, \citenamefont {Fuchs}, \citenamefont {Kimble},\ and\ \citenamefont {Polzik}}]{Furusawa1998}%
  \BibitemOpen
  \bibfield  {author} {\bibinfo {author} {\bibfnamefont {A.}~\bibnamefont {Furusawa}}, \bibinfo {author} {\bibfnamefont {J.~L.}\ \bibnamefont {Sørensen}}, \bibinfo {author} {\bibfnamefont {S.~L.}\ \bibnamefont {Braunstein}}, \bibinfo {author} {\bibfnamefont {C.~A.}\ \bibnamefont {Fuchs}}, \bibinfo {author} {\bibfnamefont {H.~J.}\ \bibnamefont {Kimble}},\ and\ \bibinfo {author} {\bibfnamefont {E.~S.}\ \bibnamefont {Polzik}},\ }\bibfield  {title} {\bibinfo {title} {Unconditional quantum teleportation},\ }\href {https://doi.org/10.1126/science.282.5389.706} {\bibfield  {journal} {\bibinfo  {journal} {Science}\ }\textbf {\bibinfo {volume} {282}},\ \bibinfo {pages} {706} (\bibinfo {year} {1998}{\natexlab{a}})},\ \Eprint {https://arxiv.org/abs/https://www.science.org/doi/pdf/10.1126/science.282.5389.706} {https://www.science.org/doi/pdf/10.1126/science.282.5389.706} \BibitemShut {NoStop}%
\bibitem [{\citenamefont {Grosshans}\ and\ \citenamefont {Grangier}(2002)}]{Grosshans2002}%
  \BibitemOpen
  \bibfield  {author} {\bibinfo {author} {\bibfnamefont {F.}~\bibnamefont {Grosshans}}\ and\ \bibinfo {author} {\bibfnamefont {P.}~\bibnamefont {Grangier}},\ }\bibfield  {title} {\bibinfo {title} {Continuous variable quantum cryptography using coherent states},\ }\href {https://doi.org/10.1103/PhysRevLett.88.057902} {\bibfield  {journal} {\bibinfo  {journal} {Phys. Rev. Lett.}\ }\textbf {\bibinfo {volume} {88}},\ \bibinfo {pages} {057902} (\bibinfo {year} {2002})}\BibitemShut {NoStop}%
\bibitem [{\citenamefont {Grosshans}\ \emph {et~al.}(2003)\citenamefont {Grosshans}, \citenamefont {Van~Assche}, \citenamefont {Wenger}, \citenamefont {Brouri}, \citenamefont {Cerf},\ and\ \citenamefont {Grangier}}]{Grosshans2003}%
  \BibitemOpen
  \bibfield  {author} {\bibinfo {author} {\bibfnamefont {F.}~\bibnamefont {Grosshans}}, \bibinfo {author} {\bibfnamefont {G.}~\bibnamefont {Van~Assche}}, \bibinfo {author} {\bibfnamefont {J.}~\bibnamefont {Wenger}}, \bibinfo {author} {\bibfnamefont {R.}~\bibnamefont {Brouri}}, \bibinfo {author} {\bibfnamefont {N.~J.}\ \bibnamefont {Cerf}},\ and\ \bibinfo {author} {\bibfnamefont {P.}~\bibnamefont {Grangier}},\ }\bibfield  {title} {\bibinfo {title} {Quantum key distribution using gaussian-modulated coherent states},\ }\href {https://doi.org/10.1038/nature01289} {\bibfield  {journal} {\bibinfo  {journal} {Nature}\ }\textbf {\bibinfo {volume} {421}},\ \bibinfo {pages} {238} (\bibinfo {year} {2003})}\BibitemShut {NoStop}%
\bibitem [{\citenamefont {Chen}\ \emph {et~al.}(2013)\citenamefont {Chen}, \citenamefont {Menicucci},\ and\ \citenamefont {Pfister}}]{Chen2013}%
  \BibitemOpen
  \bibfield  {author} {\bibinfo {author} {\bibfnamefont {M.}~\bibnamefont {Chen}}, \bibinfo {author} {\bibfnamefont {N.~C.}\ \bibnamefont {Menicucci}},\ and\ \bibinfo {author} {\bibfnamefont {O.}~\bibnamefont {Pfister}},\ }\bibfield  {title} {\bibinfo {title} {{Experimental realization of multipartite entanglement of 60 modes of a quantum optical frequency comb}},\ }\href {https://doi.org/10.1103/PhysRevLett.112.120505} {\bibfield  {journal} {\bibinfo  {journal} {Physical Review Letters}\ }\textbf {\bibinfo {volume} {112}},\ \bibinfo {pages} {1} (\bibinfo {year} {2013})},\ \Eprint {https://arxiv.org/abs/1311.2957} {arXiv:1311.2957} \BibitemShut {NoStop}%
\bibitem [{\citenamefont {Yokoyama}\ \emph {et~al.}(2013)\citenamefont {Yokoyama}, \citenamefont {Ukai}, \citenamefont {Armstrong}, \citenamefont {Sornphiphatphong}, \citenamefont {Kaji}, \citenamefont {Suzuki}, \citenamefont {Yoshikawa}, \citenamefont {Yonezawa}, \citenamefont {Menicucci},\ and\ \citenamefont {Furusawa}}]{Yokoyama2013}%
  \BibitemOpen
  \bibfield  {author} {\bibinfo {author} {\bibfnamefont {S.}~\bibnamefont {Yokoyama}}, \bibinfo {author} {\bibfnamefont {R.}~\bibnamefont {Ukai}}, \bibinfo {author} {\bibfnamefont {S.~C.}\ \bibnamefont {Armstrong}}, \bibinfo {author} {\bibfnamefont {C.}~\bibnamefont {Sornphiphatphong}}, \bibinfo {author} {\bibfnamefont {T.}~\bibnamefont {Kaji}}, \bibinfo {author} {\bibfnamefont {S.}~\bibnamefont {Suzuki}}, \bibinfo {author} {\bibfnamefont {J.-i.}\ \bibnamefont {Yoshikawa}}, \bibinfo {author} {\bibfnamefont {H.}~\bibnamefont {Yonezawa}}, \bibinfo {author} {\bibfnamefont {N.~C.}\ \bibnamefont {Menicucci}},\ and\ \bibinfo {author} {\bibfnamefont {A.}~\bibnamefont {Furusawa}},\ }\bibfield  {title} {\bibinfo {title} {Ultra-large-scale continuous-variable cluster states multiplexed in the time domain},\ }\href {https://doi.org/10.1038/nphoton.2013.287} {\bibfield  {journal} {\bibinfo  {journal} {Nature Photonics}\ }\textbf {\bibinfo {volume} {7}},\ \bibinfo {pages} {982} (\bibinfo {year} {2013})}\BibitemShut
  {NoStop}%
\bibitem [{\citenamefont {Furusawa}\ \emph {et~al.}(1998{\natexlab{b}})\citenamefont {Furusawa}, \citenamefont {Sørensen}, \citenamefont {Braunstein}, \citenamefont {Fuchs}, \citenamefont {Kimble},\ and\ \citenamefont {Polzik}}]{Furasawa1998}%
  \BibitemOpen
  \bibfield  {author} {\bibinfo {author} {\bibfnamefont {A.}~\bibnamefont {Furusawa}}, \bibinfo {author} {\bibfnamefont {J.~L.}\ \bibnamefont {Sørensen}}, \bibinfo {author} {\bibfnamefont {S.~L.}\ \bibnamefont {Braunstein}}, \bibinfo {author} {\bibfnamefont {C.~A.}\ \bibnamefont {Fuchs}}, \bibinfo {author} {\bibfnamefont {H.~J.}\ \bibnamefont {Kimble}},\ and\ \bibinfo {author} {\bibfnamefont {E.~S.}\ \bibnamefont {Polzik}},\ }\bibfield  {title} {\bibinfo {title} {Unconditional quantum teleportation},\ }\href {https://doi.org/10.1126/science.282.5389.706} {\bibfield  {journal} {\bibinfo  {journal} {Science}\ }\textbf {\bibinfo {volume} {282}},\ \bibinfo {pages} {706} (\bibinfo {year} {1998}{\natexlab{b}})},\ \Eprint {https://arxiv.org/abs/https://www.science.org/doi/pdf/10.1126/science.282.5389.706} {https://www.science.org/doi/pdf/10.1126/science.282.5389.706} \BibitemShut {NoStop}%
\bibitem [{\citenamefont {Sherson}\ \emph {et~al.}(2006)\citenamefont {Sherson}, \citenamefont {Krauter}, \citenamefont {Olsson}, \citenamefont {Julsgaard}, \citenamefont {Hammerer}, \citenamefont {Cirac},\ and\ \citenamefont {Polzik}}]{Sherson2006}%
  \BibitemOpen
  \bibfield  {author} {\bibinfo {author} {\bibfnamefont {J.~F.}\ \bibnamefont {Sherson}}, \bibinfo {author} {\bibfnamefont {H.}~\bibnamefont {Krauter}}, \bibinfo {author} {\bibfnamefont {R.~K.}\ \bibnamefont {Olsson}}, \bibinfo {author} {\bibfnamefont {B.}~\bibnamefont {Julsgaard}}, \bibinfo {author} {\bibfnamefont {K.}~\bibnamefont {Hammerer}}, \bibinfo {author} {\bibfnamefont {I.}~\bibnamefont {Cirac}},\ and\ \bibinfo {author} {\bibfnamefont {E.~S.}\ \bibnamefont {Polzik}},\ }\bibfield  {title} {\bibinfo {title} {Quantum teleportation between light and matter},\ }\href {https://doi.org/10.1038/nature05136} {\bibfield  {journal} {\bibinfo  {journal} {Nature}\ }\textbf {\bibinfo {volume} {443}},\ \bibinfo {pages} {557} (\bibinfo {year} {2006})}\BibitemShut {NoStop}%
\bibitem [{\citenamefont {Simon}\ \emph {et~al.}(1994)\citenamefont {Simon}, \citenamefont {Mukunda},\ and\ \citenamefont {Dutta}}]{Simon1994}%
  \BibitemOpen
  \bibfield  {author} {\bibinfo {author} {\bibfnamefont {R.}~\bibnamefont {Simon}}, \bibinfo {author} {\bibfnamefont {N.}~\bibnamefont {Mukunda}},\ and\ \bibinfo {author} {\bibfnamefont {B.}~\bibnamefont {Dutta}},\ }\bibfield  {title} {\bibinfo {title} {Quantum-noise matrix for multimode systems: U(n) invariance, squeezing, and normal forms},\ }\href {https://doi.org/10.1103/PhysRevA.49.1567} {\bibfield  {journal} {\bibinfo  {journal} {Phys. Rev. A}\ }\textbf {\bibinfo {volume} {49}},\ \bibinfo {pages} {1567} (\bibinfo {year} {1994})}\BibitemShut {NoStop}%
\bibitem [{\citenamefont {van Loock}\ and\ \citenamefont {Furusawa}(2003)}]{VanLoock2003}%
  \BibitemOpen
  \bibfield  {author} {\bibinfo {author} {\bibfnamefont {P.}~\bibnamefont {van Loock}}\ and\ \bibinfo {author} {\bibfnamefont {A.}~\bibnamefont {Furusawa}},\ }\bibfield  {title} {\bibinfo {title} {Detecting genuine multipartite continuous-variable entanglement},\ }\href {https://doi.org/10.1103/PhysRevA.67.052315} {\bibfield  {journal} {\bibinfo  {journal} {Phys. Rev. A}\ }\textbf {\bibinfo {volume} {67}},\ \bibinfo {pages} {052315} (\bibinfo {year} {2003})}\BibitemShut {NoStop}%
\bibitem [{\citenamefont {Martinelli}(2023)}]{Martinelli2023}%
  \BibitemOpen
  \bibfield  {author} {\bibinfo {author} {\bibfnamefont {M.}~\bibnamefont {Martinelli}},\ }\bibfield  {title} {\bibinfo {title} {From spectral matrix to sideband structure: exploring stereo multimodes},\ }\href {https://doi.org/10.1364/JOSAB.482651} {\bibfield  {journal} {\bibinfo  {journal} {J. Opt. Soc. Am. B}\ }\textbf {\bibinfo {volume} {40}},\ \bibinfo {pages} {1277} (\bibinfo {year} {2023})}\BibitemShut {NoStop}%
\bibitem [{\citenamefont {Zhang}\ and\ \citenamefont {M\o{}lmer}(2017)}]{Zhang2017}%
  \BibitemOpen
  \bibfield  {author} {\bibinfo {author} {\bibfnamefont {J.}~\bibnamefont {Zhang}}\ and\ \bibinfo {author} {\bibfnamefont {K.}~\bibnamefont {M\o{}lmer}},\ }\bibfield  {title} {\bibinfo {title} {Prediction and retrodiction with continuously monitored gaussian states},\ }\href {https://doi.org/10.1103/PhysRevA.96.062131} {\bibfield  {journal} {\bibinfo  {journal} {Phys. Rev. A}\ }\textbf {\bibinfo {volume} {96}},\ \bibinfo {pages} {062131} (\bibinfo {year} {2017})}\BibitemShut {NoStop}%
\bibitem [{\citenamefont {Brandão}\ \emph {et~al.}(2022)\citenamefont {Brandão}, \citenamefont {Tandeitnik},\ and\ \citenamefont {Guerreiro}}]{Brandao2022}%
  \BibitemOpen
  \bibfield  {author} {\bibinfo {author} {\bibfnamefont {I.}~\bibnamefont {Brandão}}, \bibinfo {author} {\bibfnamefont {D.}~\bibnamefont {Tandeitnik}},\ and\ \bibinfo {author} {\bibfnamefont {T.}~\bibnamefont {Guerreiro}},\ }\bibfield  {title} {\bibinfo {title} {Qugit: A numerical toolbox for gaussian quantum states},\ }\href {https://doi.org/https://doi.org/10.1016/j.cpc.2022.108471} {\bibfield  {journal} {\bibinfo  {journal} {Computer Physics Communications}\ }\textbf {\bibinfo {volume} {280}},\ \bibinfo {pages} {108471} (\bibinfo {year} {2022})}\BibitemShut {NoStop}%
\bibitem [{\citenamefont {Kolobov}\ and\ \citenamefont {Patera}(2011)}]{Kolobov2011}%
  \BibitemOpen
  \bibfield  {author} {\bibinfo {author} {\bibfnamefont {M.~I.}\ \bibnamefont {Kolobov}}\ and\ \bibinfo {author} {\bibfnamefont {G.}~\bibnamefont {Patera}},\ }\bibfield  {title} {\bibinfo {title} {Spatiotemporal multipartite entanglement},\ }\href {https://doi.org/10.1103/PhysRevA.83.050302} {\bibfield  {journal} {\bibinfo  {journal} {Phys. Rev. A}\ }\textbf {\bibinfo {volume} {83}},\ \bibinfo {pages} {050302} (\bibinfo {year} {2011})}\BibitemShut {NoStop}%
\bibitem [{\citenamefont {Chembo}(2016)}]{Chembo2016}%
  \BibitemOpen
  \bibfield  {author} {\bibinfo {author} {\bibfnamefont {Y.~K.}\ \bibnamefont {Chembo}},\ }\bibfield  {title} {\bibinfo {title} {{Quantum dynamics of Kerr optical frequency combs below and above threshold: Spontaneous four-wave mixing, entanglement, and squeezed states of light}},\ }\href {https://doi.org/10.1103/PhysRevA.93.033820} {\bibfield  {journal} {\bibinfo  {journal} {Physical Review A}\ }\textbf {\bibinfo {volume} {93}},\ \bibinfo {pages} {033820} (\bibinfo {year} {2016})}\BibitemShut {NoStop}%
\bibitem [{\citenamefont {D'Auria}\ \emph {et~al.}(2009)\citenamefont {D'Auria}, \citenamefont {Fornaro}, \citenamefont {Porzio}, \citenamefont {Solimeno}, \citenamefont {Olivares},\ and\ \citenamefont {Paris}}]{DAuria2009}%
  \BibitemOpen
  \bibfield  {author} {\bibinfo {author} {\bibfnamefont {V.}~\bibnamefont {D'Auria}}, \bibinfo {author} {\bibfnamefont {S.}~\bibnamefont {Fornaro}}, \bibinfo {author} {\bibfnamefont {A.}~\bibnamefont {Porzio}}, \bibinfo {author} {\bibfnamefont {S.}~\bibnamefont {Solimeno}}, \bibinfo {author} {\bibfnamefont {S.}~\bibnamefont {Olivares}},\ and\ \bibinfo {author} {\bibfnamefont {M.~G.~A.}\ \bibnamefont {Paris}},\ }\bibfield  {title} {\bibinfo {title} {Full characterization of gaussian bipartite entangled states by a single homodyne detector},\ }\href {https://doi.org/10.1103/PhysRevLett.102.020502} {\bibfield  {journal} {\bibinfo  {journal} {Phys. Rev. Lett.}\ }\textbf {\bibinfo {volume} {102}},\ \bibinfo {pages} {020502} (\bibinfo {year} {2009})}\BibitemShut {NoStop}%
\bibitem [{\citenamefont {Laurat}\ \emph {et~al.}(2005)\citenamefont {Laurat}, \citenamefont {Longchambon}, \citenamefont {Fabre},\ and\ \citenamefont {Coudreau}}]{Laurat2005}%
  \BibitemOpen
  \bibfield  {author} {\bibinfo {author} {\bibfnamefont {J.}~\bibnamefont {Laurat}}, \bibinfo {author} {\bibfnamefont {L.}~\bibnamefont {Longchambon}}, \bibinfo {author} {\bibfnamefont {C.}~\bibnamefont {Fabre}},\ and\ \bibinfo {author} {\bibfnamefont {T.}~\bibnamefont {Coudreau}},\ }\bibfield  {title} {\bibinfo {title} {Experimental investigation of amplitude and phase quantum correlations in a type ii optical parametric oscillator above threshold: from nondegenerate to degenerate operation},\ }\href {https://doi.org/10.1364/OL.30.001177} {\bibfield  {journal} {\bibinfo  {journal} {Opt. Lett.}\ }\textbf {\bibinfo {volume} {30}},\ \bibinfo {pages} {1177} (\bibinfo {year} {2005})}\BibitemShut {NoStop}%
\bibitem [{\citenamefont {Guidry}\ \emph {et~al.}(2023)\citenamefont {Guidry}, \citenamefont {Lukin}, \citenamefont {Yang},\ and\ \citenamefont {Vu\v{c}kovi\'{c}}}]{Guidry2023}%
  \BibitemOpen
  \bibfield  {author} {\bibinfo {author} {\bibfnamefont {M.~A.}\ \bibnamefont {Guidry}}, \bibinfo {author} {\bibfnamefont {D.~M.}\ \bibnamefont {Lukin}}, \bibinfo {author} {\bibfnamefont {K.~Y.}\ \bibnamefont {Yang}},\ and\ \bibinfo {author} {\bibfnamefont {J.}~\bibnamefont {Vu\v{c}kovi\'{c}}},\ }\bibfield  {title} {\bibinfo {title} {Multimode squeezing in soliton crystal microcombs},\ }\href {https://doi.org/10.1364/OPTICA.485996} {\bibfield  {journal} {\bibinfo  {journal} {Optica}\ }\textbf {\bibinfo {volume} {10}},\ \bibinfo {pages} {694} (\bibinfo {year} {2023})}\BibitemShut {NoStop}%
\bibitem [{\citenamefont {Meng}\ \emph {et~al.}(2020)\citenamefont {Meng}, \citenamefont {Brawley}, \citenamefont {Bennett}, \citenamefont {Vanner},\ and\ \citenamefont {Bowen}}]{Meng2020}%
  \BibitemOpen
  \bibfield  {author} {\bibinfo {author} {\bibfnamefont {C.}~\bibnamefont {Meng}}, \bibinfo {author} {\bibfnamefont {G.~A.}\ \bibnamefont {Brawley}}, \bibinfo {author} {\bibfnamefont {J.~S.}\ \bibnamefont {Bennett}}, \bibinfo {author} {\bibfnamefont {M.~R.}\ \bibnamefont {Vanner}},\ and\ \bibinfo {author} {\bibfnamefont {W.~P.}\ \bibnamefont {Bowen}},\ }\bibfield  {title} {\bibinfo {title} {Mechanical squeezing via fast continuous measurement},\ }\href {https://doi.org/10.1103/PhysRevLett.125.043604} {\bibfield  {journal} {\bibinfo  {journal} {Phys. Rev. Lett.}\ }\textbf {\bibinfo {volume} {125}},\ \bibinfo {pages} {043604} (\bibinfo {year} {2020})}\BibitemShut {NoStop}%
\bibitem [{\citenamefont {Isaksen}\ and\ \citenamefont {Andersen}(2023)}]{Isaksen2023}%
  \BibitemOpen
  \bibfield  {author} {\bibinfo {author} {\bibfnamefont {F.~W.}\ \bibnamefont {Isaksen}}\ and\ \bibinfo {author} {\bibfnamefont {U.~L.}\ \bibnamefont {Andersen}},\ }\bibfield  {title} {\bibinfo {title} {Mechanical cooling and squeezing using optimal control},\ }\href {https://doi.org/10.1103/PhysRevA.107.023512} {\bibfield  {journal} {\bibinfo  {journal} {Phys. Rev. A}\ }\textbf {\bibinfo {volume} {107}},\ \bibinfo {pages} {023512} (\bibinfo {year} {2023})}\BibitemShut {NoStop}%
\bibitem [{\citenamefont {Gouzien}\ \emph {et~al.}(2023)\citenamefont {Gouzien}, \citenamefont {Labont{\'{e}}}, \citenamefont {Etesse}, \citenamefont {Zavatta}, \citenamefont {Tanzilli}, \citenamefont {D'Auria},\ and\ \citenamefont {Patera}}]{Gouzien2023}%
  \BibitemOpen
  \bibfield  {author} {\bibinfo {author} {\bibfnamefont {{\'{E}}.}~\bibnamefont {Gouzien}}, \bibinfo {author} {\bibfnamefont {L.}~\bibnamefont {Labont{\'{e}}}}, \bibinfo {author} {\bibfnamefont {J.}~\bibnamefont {Etesse}}, \bibinfo {author} {\bibfnamefont {A.}~\bibnamefont {Zavatta}}, \bibinfo {author} {\bibfnamefont {S.}~\bibnamefont {Tanzilli}}, \bibinfo {author} {\bibfnamefont {V.}~\bibnamefont {D'Auria}},\ and\ \bibinfo {author} {\bibfnamefont {G.}~\bibnamefont {Patera}},\ }\bibfield  {title} {\bibinfo {title} {{Hidden and detectable squeezing from microresonators}},\ }\href {https://doi.org/10.1103/physrevresearch.5.023178} {\bibfield  {journal} {\bibinfo  {journal} {Physical Review Research}\ }\textbf {\bibinfo {volume} {5}},\ \bibinfo {pages} {1} (\bibinfo {year} {2023})},\ \Eprint {https://arxiv.org/abs/2207.00360} {arXiv:2207.00360} \BibitemShut {NoStop}%
\bibitem [{\citenamefont {Lvovsky}\ and\ \citenamefont {Raymer}(2009)}]{Lvovsky2009a}%
  \BibitemOpen
  \bibfield  {author} {\bibinfo {author} {\bibfnamefont {A.~I.}\ \bibnamefont {Lvovsky}}\ and\ \bibinfo {author} {\bibfnamefont {M.~G.}\ \bibnamefont {Raymer}},\ }\bibfield  {title} {\bibinfo {title} {{Continuous-variable optical quantum-state tomography}},\ }\href {https://doi.org/10.1103/RevModPhys.81.299} {\bibfield  {journal} {\bibinfo  {journal} {Reviews of Modern Physics}\ }\textbf {\bibinfo {volume} {81}},\ \bibinfo {pages} {299} (\bibinfo {year} {2009})},\ \Eprint {https://arxiv.org/abs/0511044} {arXiv:0511044 [quant-ph]} \BibitemShut {NoStop}%
\bibitem [{\citenamefont {Yurke}(1985)}]{Yurke1985}%
  \BibitemOpen
  \bibfield  {author} {\bibinfo {author} {\bibfnamefont {B.}~\bibnamefont {Yurke}},\ }\bibfield  {title} {\bibinfo {title} {{Squeezed-coherent-state generation via four-wave mixers and detection via homodyne detectors}},\ }\href {https://doi.org/10.1103/PhysRevA.32.300} {\bibfield  {journal} {\bibinfo  {journal} {Physical Review A}\ }\textbf {\bibinfo {volume} {32}},\ \bibinfo {pages} {300} (\bibinfo {year} {1985})}\BibitemShut {NoStop}%
\bibitem [{\citenamefont {Gouzien}\ \emph {et~al.}(2020)\citenamefont {Gouzien}, \citenamefont {Tanzilli}, \citenamefont {D'Auria},\ and\ \citenamefont {Patera}}]{Gouzien2020}%
  \BibitemOpen
  \bibfield  {author} {\bibinfo {author} {\bibfnamefont {{\'{E}}.}~\bibnamefont {Gouzien}}, \bibinfo {author} {\bibfnamefont {S.}~\bibnamefont {Tanzilli}}, \bibinfo {author} {\bibfnamefont {V.}~\bibnamefont {D'Auria}},\ and\ \bibinfo {author} {\bibfnamefont {G.}~\bibnamefont {Patera}},\ }\bibfield  {title} {\bibinfo {title} {{Morphing Supermodes: A Full Characterization for Enabling Multimode Quantum Optics}},\ }\href {https://doi.org/10.1103/PhysRevLett.125.103601} {\bibfield  {journal} {\bibinfo  {journal} {Physical Review Letters}\ }\textbf {\bibinfo {volume} {125}},\ \bibinfo {pages} {1} (\bibinfo {year} {2020})}\BibitemShut {NoStop}%
\bibitem [{\citenamefont {Barbosa}\ \emph {et~al.}(2013{\natexlab{a}})\citenamefont {Barbosa}, \citenamefont {Coelho}, \citenamefont {Cassemiro}, \citenamefont {Nussenzveig}, \citenamefont {Fabre}, \citenamefont {Martinelli},\ and\ \citenamefont {Villar}}]{Barbosa2013a}%
  \BibitemOpen
  \bibfield  {author} {\bibinfo {author} {\bibfnamefont {F.~A.}\ \bibnamefont {Barbosa}}, \bibinfo {author} {\bibfnamefont {A.~S.}\ \bibnamefont {Coelho}}, \bibinfo {author} {\bibfnamefont {K.~N.}\ \bibnamefont {Cassemiro}}, \bibinfo {author} {\bibfnamefont {P.}~\bibnamefont {Nussenzveig}}, \bibinfo {author} {\bibfnamefont {C.}~\bibnamefont {Fabre}}, \bibinfo {author} {\bibfnamefont {M.}~\bibnamefont {Martinelli}},\ and\ \bibinfo {author} {\bibfnamefont {A.~S.}\ \bibnamefont {Villar}},\ }\bibfield  {title} {\bibinfo {title} {{Beyond spectral homodyne detection: Complete quantum measurement of spectral modes of light}},\ }\href {https://doi.org/10.1103/PhysRevLett.111.200402} {\bibfield  {journal} {\bibinfo  {journal} {Physical Review Letters}\ }\textbf {\bibinfo {volume} {111}},\ \bibinfo {pages} {1} (\bibinfo {year} {2013}{\natexlab{a}})},\ \Eprint {https://arxiv.org/abs/1308.5650} {arXiv:1308.5650} \BibitemShut {NoStop}%
\bibitem [{\citenamefont {Barbosa}\ \emph {et~al.}(2013{\natexlab{b}})\citenamefont {Barbosa}, \citenamefont {Coelho}, \citenamefont {Cassemiro}, \citenamefont {Nussenzveig}, \citenamefont {Fabre}, \citenamefont {Villar},\ and\ \citenamefont {Martinelli}}]{Barbosa2013b}%
  \BibitemOpen
  \bibfield  {author} {\bibinfo {author} {\bibfnamefont {F.~A.~S.}\ \bibnamefont {Barbosa}}, \bibinfo {author} {\bibfnamefont {A.~S.}\ \bibnamefont {Coelho}}, \bibinfo {author} {\bibfnamefont {K.~N.}\ \bibnamefont {Cassemiro}}, \bibinfo {author} {\bibfnamefont {P.}~\bibnamefont {Nussenzveig}}, \bibinfo {author} {\bibfnamefont {C.}~\bibnamefont {Fabre}}, \bibinfo {author} {\bibfnamefont {A.~S.}\ \bibnamefont {Villar}},\ and\ \bibinfo {author} {\bibfnamefont {M.}~\bibnamefont {Martinelli}},\ }\bibfield  {title} {\bibinfo {title} {Quantum state reconstruction of spectral field modes: Homodyne and resonator detection schemes},\ }\href {https://doi.org/10.1103/PhysRevA.88.052113} {\bibfield  {journal} {\bibinfo  {journal} {Phys. Rev. A}\ }\textbf {\bibinfo {volume} {88}},\ \bibinfo {pages} {052113} (\bibinfo {year} {2013}{\natexlab{b}})}\BibitemShut {NoStop}%
\bibitem [{\citenamefont {Buchmann}\ \emph {et~al.}(2016)\citenamefont {Buchmann}, \citenamefont {Schreppler}, \citenamefont {Kohler}, \citenamefont {Spethmann},\ and\ \citenamefont {Stamper-Kurn}}]{Buchmann2016}%
  \BibitemOpen
  \bibfield  {author} {\bibinfo {author} {\bibfnamefont {L.~F.}\ \bibnamefont {Buchmann}}, \bibinfo {author} {\bibfnamefont {S.}~\bibnamefont {Schreppler}}, \bibinfo {author} {\bibfnamefont {J.}~\bibnamefont {Kohler}}, \bibinfo {author} {\bibfnamefont {N.}~\bibnamefont {Spethmann}},\ and\ \bibinfo {author} {\bibfnamefont {D.~M.}\ \bibnamefont {Stamper-Kurn}},\ }\bibfield  {title} {\bibinfo {title} {{Complex Squeezing and Force Measurement beyond the Standard Quantum Limit}},\ }\href {https://doi.org/10.1103/PhysRevLett.117.030801} {\bibfield  {journal} {\bibinfo  {journal} {Physical Review Letters}\ }\textbf {\bibinfo {volume} {117}},\ \bibinfo {pages} {1} (\bibinfo {year} {2016})},\ \Eprint {https://arxiv.org/abs/1602.02141} {arXiv:1602.02141} \BibitemShut {NoStop}%
\bibitem [{\citenamefont {Lvovsky}\ \emph {et~al.}(2020)\citenamefont {Lvovsky}, \citenamefont {Grangier}, \citenamefont {Ourjoumtsev}, \citenamefont {Parigi}, \citenamefont {Sasaki},\ and\ \citenamefont {Tualle-Brouri}}]{Lvovsky2020}%
  \BibitemOpen
  \bibfield  {author} {\bibinfo {author} {\bibfnamefont {A.~I.}\ \bibnamefont {Lvovsky}}, \bibinfo {author} {\bibfnamefont {P.}~\bibnamefont {Grangier}}, \bibinfo {author} {\bibfnamefont {A.}~\bibnamefont {Ourjoumtsev}}, \bibinfo {author} {\bibfnamefont {V.}~\bibnamefont {Parigi}}, \bibinfo {author} {\bibfnamefont {M.}~\bibnamefont {Sasaki}},\ and\ \bibinfo {author} {\bibfnamefont {R.}~\bibnamefont {Tualle-Brouri}},\ }\href {https://arxiv.org/abs/2006.16985} {\bibinfo {title} {Production and applications of non-gaussian quantum states of light}} (\bibinfo {year} {2020}),\ \Eprint {https://arxiv.org/abs/2006.16985} {arXiv:2006.16985 [quant-ph]} \BibitemShut {NoStop}%
\bibitem [{\citenamefont {Dioum}\ \emph {et~al.}(2024)\citenamefont {Dioum}, \citenamefont {D'Auria}, \citenamefont {Zavatta}, \citenamefont {Pfister},\ and\ \citenamefont {Patera}}]{Dioum2024}%
  \BibitemOpen
  \bibfield  {author} {\bibinfo {author} {\bibfnamefont {B.}~\bibnamefont {Dioum}}, \bibinfo {author} {\bibfnamefont {V.}~\bibnamefont {D'Auria}}, \bibinfo {author} {\bibfnamefont {A.}~\bibnamefont {Zavatta}}, \bibinfo {author} {\bibfnamefont {O.}~\bibnamefont {Pfister}},\ and\ \bibinfo {author} {\bibfnamefont {G.}~\bibnamefont {Patera}},\ }\href@noop {} {\bibinfo {title} {Universal quantum frequency comb measurements by spectral mode-matching}} (\bibinfo {year} {2024}),\ \Eprint {https://arxiv.org/abs/2405.18454} {arXiv:2405.18454 [quant-ph]} \BibitemShut {NoStop}%
\bibitem [{\citenamefont {Vaidya}\ \emph {et~al.}(2020)\citenamefont {Vaidya}, \citenamefont {Morrison}, \citenamefont {Helt}, \citenamefont {Shahrokshahi}, \citenamefont {Mahler}, \citenamefont {Collins}, \citenamefont {Tan}, \citenamefont {Lavoie}, \citenamefont {Repingon}, \citenamefont {Menotti}, \citenamefont {Quesada}, \citenamefont {Pooser}, \citenamefont {Lita}, \citenamefont {Gerrits}, \citenamefont {Nam},\ and\ \citenamefont {Vernon}}]{Vaidya2020a}%
  \BibitemOpen
  \bibfield  {author} {\bibinfo {author} {\bibfnamefont {V.~D.}\ \bibnamefont {Vaidya}}, \bibinfo {author} {\bibfnamefont {B.}~\bibnamefont {Morrison}}, \bibinfo {author} {\bibfnamefont {L.~G.}\ \bibnamefont {Helt}}, \bibinfo {author} {\bibfnamefont {R.}~\bibnamefont {Shahrokshahi}}, \bibinfo {author} {\bibfnamefont {D.~H.}\ \bibnamefont {Mahler}}, \bibinfo {author} {\bibfnamefont {M.~J.}\ \bibnamefont {Collins}}, \bibinfo {author} {\bibfnamefont {K.}~\bibnamefont {Tan}}, \bibinfo {author} {\bibfnamefont {J.}~\bibnamefont {Lavoie}}, \bibinfo {author} {\bibfnamefont {A.}~\bibnamefont {Repingon}}, \bibinfo {author} {\bibfnamefont {M.}~\bibnamefont {Menotti}}, \bibinfo {author} {\bibfnamefont {N.}~\bibnamefont {Quesada}}, \bibinfo {author} {\bibfnamefont {R.~C.}\ \bibnamefont {Pooser}}, \bibinfo {author} {\bibfnamefont {A.~E.}\ \bibnamefont {Lita}}, \bibinfo {author} {\bibfnamefont {T.}~\bibnamefont {Gerrits}}, \bibinfo {author} {\bibfnamefont {S.~W.}\ \bibnamefont {Nam}},\ and\ \bibinfo {author} {\bibfnamefont
  {Z.}~\bibnamefont {Vernon}},\ }\bibfield  {title} {\bibinfo {title} {Broadband quadrature-squeezed vacuum and nonclassical photon number correlations from a nanophotonic device},\ }\href {https://doi.org/10.1126/sciadv.aba9186} {\bibfield  {journal} {\bibinfo  {journal} {Science Advances}\ }\textbf {\bibinfo {volume} {6}},\ \bibinfo {pages} {eaba9186} (\bibinfo {year} {2020})}\BibitemShut {NoStop}%
\bibitem [{\citenamefont {Seifoory}\ \emph {et~al.}(2022)\citenamefont {Seifoory}, \citenamefont {Vernon}, \citenamefont {Mahler}, \citenamefont {Menotti}, \citenamefont {Zhang},\ and\ \citenamefont {Sipe}}]{Seifoory2022}%
  \BibitemOpen
  \bibfield  {author} {\bibinfo {author} {\bibfnamefont {H.}~\bibnamefont {Seifoory}}, \bibinfo {author} {\bibfnamefont {Z.}~\bibnamefont {Vernon}}, \bibinfo {author} {\bibfnamefont {D.~H.}\ \bibnamefont {Mahler}}, \bibinfo {author} {\bibfnamefont {M.}~\bibnamefont {Menotti}}, \bibinfo {author} {\bibfnamefont {Y.}~\bibnamefont {Zhang}},\ and\ \bibinfo {author} {\bibfnamefont {J.~E.}\ \bibnamefont {Sipe}},\ }\bibfield  {title} {\bibinfo {title} {{Degenerate squeezing in a dual-pumped integrated microresonator: Parasitic processes and their suppression}},\ }\href {https://doi.org/10.1103/physreva.105.033524} {\bibfield  {journal} {\bibinfo  {journal} {Physical Review A}\ }\textbf {\bibinfo {volume} {105}},\ \bibinfo {pages} {1} (\bibinfo {year} {2022})},\ \Eprint {https://arxiv.org/abs/2109.03298} {arXiv:2109.03298} \BibitemShut {NoStop}%
\bibitem [{Sym()}]{SympForm}%
  \BibitemOpen
  \href@noop {} {}\bibinfo {note} {$\Omega$ is the $2N \times 2N$ symplectic form defined as $\Omega=\left(\begin{array}{c|c}0 & I_{\mathrm{N}}\\ \hline-I_{\mathrm{N}} & 0\end{array}\right)$ where $I_{\mathrm{N}}$ is the identity matrix of dimension $N$.}\BibitemShut {Stop}%
\bibitem [{\citenamefont {Lvovsky}(2016)}]{Lvovsky2016}%
  \BibitemOpen
  \bibfield  {author} {\bibinfo {author} {\bibfnamefont {A.~I.}\ \bibnamefont {Lvovsky}},\ }\href {https://arxiv.org/abs/1401.4118} {\bibinfo {title} {Squeezed light}} (\bibinfo {year} {2016}),\ \Eprint {https://arxiv.org/abs/1401.4118} {arXiv:1401.4118 [quant-ph]} \BibitemShut {NoStop}%
\bibitem [{\citenamefont {Barbosa}\ \emph {et~al.}(2018)\citenamefont {Barbosa}, \citenamefont {Coelho}, \citenamefont {Mu{\~{n}}oz-Mart{\'{i}}nez}, \citenamefont {Ortiz-Guti{\'{e}}rrez}, \citenamefont {Villar}, \citenamefont {Nussenzveig},\ and\ \citenamefont {Martinelli}}]{Barbosa2018a}%
  \BibitemOpen
  \bibfield  {author} {\bibinfo {author} {\bibfnamefont {F.~A.}\ \bibnamefont {Barbosa}}, \bibinfo {author} {\bibfnamefont {A.~S.}\ \bibnamefont {Coelho}}, \bibinfo {author} {\bibfnamefont {L.~F.}\ \bibnamefont {Mu{\~{n}}oz-Mart{\'{i}}nez}}, \bibinfo {author} {\bibfnamefont {L.}~\bibnamefont {Ortiz-Guti{\'{e}}rrez}}, \bibinfo {author} {\bibfnamefont {A.~S.}\ \bibnamefont {Villar}}, \bibinfo {author} {\bibfnamefont {P.}~\bibnamefont {Nussenzveig}},\ and\ \bibinfo {author} {\bibfnamefont {M.}~\bibnamefont {Martinelli}},\ }\bibfield  {title} {\bibinfo {title} {{Hexapartite Entanglement in an above-Threshold Optical Parametric Oscillator}},\ }\href {https://doi.org/10.1103/PhysRevLett.121.073601} {\bibfield  {journal} {\bibinfo  {journal} {Physical Review Letters}\ }\textbf {\bibinfo {volume} {121}},\ \bibinfo {pages} {4} (\bibinfo {year} {2018})},\ \Eprint {https://arxiv.org/abs/1712.01756} {arXiv:1712.01756} \BibitemShut {NoStop}%
\bibitem [{\citenamefont {Gardiner}\ and\ \citenamefont {Zoller}(2000)}]{gardiner2004}%
  \BibitemOpen
  \bibfield  {author} {\bibinfo {author} {\bibfnamefont {C.}~\bibnamefont {Gardiner}}\ and\ \bibinfo {author} {\bibfnamefont {P.}~\bibnamefont {Zoller}},\ }\href {https://books.google.fr/books?id=4bJ6MgEACAAJ} {\emph {\bibinfo {title} {Quantum Noise: A Handbook of Markovian and Non-Markovian Quantum Stochastic Methods with Applications to Quantum Optics}}},\ Springer series in synergetics\ (\bibinfo  {publisher} {Springer},\ \bibinfo {year} {2000})\BibitemShut {NoStop}%
\bibitem [{Note1()}]{Note1}%
  \BibitemOpen
  \bibinfo {note} {The conjugate-symplectic group Sp$^*(2N,\protect \mathbb {C})$ is defined as the set of $2N\times 2N$ complex matrices such that $S\Omega S^\dagger $, with $\Omega $ the symplectic form.}\BibitemShut {Stop}%
\bibitem [{\citenamefont {Fabre}\ and\ \citenamefont {Treps}(2020)}]{Fabre2020}%
  \BibitemOpen
  \bibfield  {author} {\bibinfo {author} {\bibfnamefont {C.}~\bibnamefont {Fabre}}\ and\ \bibinfo {author} {\bibfnamefont {N.}~\bibnamefont {Treps}},\ }\bibfield  {title} {\bibinfo {title} {{Modes and states in quantum optics}},\ }\href {https://doi.org/10.1103/REVMODPHYS.92.035005} {\bibfield  {journal} {\bibinfo  {journal} {Reviews of Modern Physics}\ }\textbf {\bibinfo {volume} {92}},\ \bibinfo {pages} {35005} (\bibinfo {year} {2020})},\ \Eprint {https://arxiv.org/abs/1912.09321} {arXiv:1912.09321} \BibitemShut {NoStop}%
\bibitem [{\citenamefont {Patera}\ \emph {et~al.}(2012)\citenamefont {Patera}, \citenamefont {Navarrete-Benlloch}, \citenamefont {de~Valc{\'a}rcel},\ and\ \citenamefont {Fabre}}]{Patera2012}%
  \BibitemOpen
  \bibfield  {author} {\bibinfo {author} {\bibfnamefont {G.}~\bibnamefont {Patera}}, \bibinfo {author} {\bibfnamefont {C.}~\bibnamefont {Navarrete-Benlloch}}, \bibinfo {author} {\bibfnamefont {G.~J.}\ \bibnamefont {de~Valc{\'a}rcel}},\ and\ \bibinfo {author} {\bibfnamefont {C.}~\bibnamefont {Fabre}},\ }\bibfield  {title} {\bibinfo {title} {Quantum coherent control of highly multipartite continuous-variable entangled states by tailoring parametric interactions},\ }\href {https://doi.org/10.1140/epjd/e2012-30036-2} {\bibfield  {journal} {\bibinfo  {journal} {The European Physical Journal D}\ }\textbf {\bibinfo {volume} {66}},\ \bibinfo {pages} {241} (\bibinfo {year} {2012})}\BibitemShut {NoStop}%
\bibitem [{\citenamefont {Arzani}\ \emph {et~al.}(2018)\citenamefont {Arzani}, \citenamefont {Fabre},\ and\ \citenamefont {Treps}}]{Arzani2018}%
  \BibitemOpen
  \bibfield  {author} {\bibinfo {author} {\bibfnamefont {F.}~\bibnamefont {Arzani}}, \bibinfo {author} {\bibfnamefont {C.}~\bibnamefont {Fabre}},\ and\ \bibinfo {author} {\bibfnamefont {N.}~\bibnamefont {Treps}},\ }\bibfield  {title} {\bibinfo {title} {Versatile engineering of multimode squeezed states by optimizing the pump spectral profile in spontaneous parametric down-conversion},\ }\href {https://doi.org/10.1103/PhysRevA.97.033808} {\bibfield  {journal} {\bibinfo  {journal} {Phys. Rev. A}\ }\textbf {\bibinfo {volume} {97}},\ \bibinfo {pages} {033808} (\bibinfo {year} {2018})}\BibitemShut {NoStop}%
\bibitem [{\citenamefont {Dieci}\ and\ \citenamefont {Pugliese}(2024)}]{Dieci2024}%
  \BibitemOpen
  \bibfield  {author} {\bibinfo {author} {\bibfnamefont {L.}~\bibnamefont {Dieci}}\ and\ \bibinfo {author} {\bibfnamefont {A.}~\bibnamefont {Pugliese}},\ }\bibfield  {title} {\bibinfo {title} {Svd, joint-mvd, berry phase, and generic loss of rank for a matrix valued function of 2 parameters},\ }\href {https://doi.org/https://doi.org/10.1016/j.laa.2024.07.021} {\bibfield  {journal} {\bibinfo  {journal} {Linear Algebra and its Applications}\ }\textbf {\bibinfo {volume} {700}},\ \bibinfo {pages} {137} (\bibinfo {year} {2024})}\BibitemShut {NoStop}%
\bibitem [{\citenamefont {Pugliese}(2024)}]{Pugliese_matlab}%
  \BibitemOpen
  \bibfield  {author} {\bibinfo {author} {\bibfnamefont {A.}~\bibnamefont {Pugliese}},\ }\href {https://fr.mathworks.com/matlabcentral/fileexchange/160876-smooth-singular-value-decomp-of-complex-matrix-function} {\bibinfo {title} {Smooth singular value decomposition of complex matrix function}},\ \bibinfo {howpublished} {MATLAB Central File Exchange} (\bibinfo {year} {2024})\BibitemShut {NoStop}%
\bibitem [{\citenamefont {Fabre}\ \emph {et~al.}(1990)\citenamefont {Fabre}, \citenamefont {Giacobino}, \citenamefont {Heidmann}, \citenamefont {Lugiato}, \citenamefont {Reynaud}, \citenamefont {Vadacchino},\ and\ \citenamefont {Kaige}}]{Fabre1990}%
  \BibitemOpen
  \bibfield  {author} {\bibinfo {author} {\bibfnamefont {C.}~\bibnamefont {Fabre}}, \bibinfo {author} {\bibfnamefont {E.}~\bibnamefont {Giacobino}}, \bibinfo {author} {\bibfnamefont {A.}~\bibnamefont {Heidmann}}, \bibinfo {author} {\bibfnamefont {L.}~\bibnamefont {Lugiato}}, \bibinfo {author} {\bibfnamefont {S.}~\bibnamefont {Reynaud}}, \bibinfo {author} {\bibfnamefont {M.}~\bibnamefont {Vadacchino}},\ and\ \bibinfo {author} {\bibfnamefont {W.}~\bibnamefont {Kaige}},\ }\bibfield  {title} {\bibinfo {title} {{Squeezing in detuned degenerate optical parametric oscillators}},\ }\href {https://doi.org/10.1088/0954-8998/2/2/006} {\bibfield  {journal} {\bibinfo  {journal} {Quantum Optics: Journal of the European Optical Society Part B}\ }\textbf {\bibinfo {volume} {2}},\ \bibinfo {pages} {159} (\bibinfo {year} {1990})}\BibitemShut {NoStop}%
\bibitem [{\citenamefont {Fabre}\ \emph {et~al.}(1989)\citenamefont {Fabre}, \citenamefont {Giacobino}, \citenamefont {Heidmann},\ and\ \citenamefont {Reynaud}}]{Fabre1989}%
  \BibitemOpen
  \bibfield  {author} {\bibinfo {author} {\bibfnamefont {C.}~\bibnamefont {Fabre}}, \bibinfo {author} {\bibfnamefont {E.}~\bibnamefont {Giacobino}}, \bibinfo {author} {\bibfnamefont {A.}~\bibnamefont {Heidmann}},\ and\ \bibinfo {author} {\bibfnamefont {S.}~\bibnamefont {Reynaud}},\ }\bibfield  {title} {\bibinfo {title} {{Noise characteristics of a non-degenerate Optical Parametric Oscillator - Application to quantum noise reduction}},\ }\href {https://doi.org/10.1051/jphys:0198900500100120900} {\bibfield  {journal} {\bibinfo  {journal} {Journal de Physique}\ }\textbf {\bibinfo {volume} {50}},\ \bibinfo {pages} {1209} (\bibinfo {year} {1989})}\BibitemShut {NoStop}%
\bibitem [{\citenamefont {Christ}\ \emph {et~al.}(2013)\citenamefont {Christ}, \citenamefont {Brecht}, \citenamefont {Mauerer},\ and\ \citenamefont {Silberhorn}}]{Christ2013c}%
  \BibitemOpen
  \bibfield  {author} {\bibinfo {author} {\bibfnamefont {A.}~\bibnamefont {Christ}}, \bibinfo {author} {\bibfnamefont {B.}~\bibnamefont {Brecht}}, \bibinfo {author} {\bibfnamefont {W.}~\bibnamefont {Mauerer}},\ and\ \bibinfo {author} {\bibfnamefont {C.}~\bibnamefont {Silberhorn}},\ }\bibfield  {title} {\bibinfo {title} {{Theory of quantum frequency conversion and type-II parametric down-conversion in the high-gain regime}},\ }\bibfield  {journal} {\bibinfo  {journal} {New Journal of Physics}\ }\textbf {\bibinfo {volume} {15}},\ \href {https://doi.org/10.1088/1367-2630/15/5/053038} {10.1088/1367-2630/15/5/053038} (\bibinfo {year} {2013}),\ \Eprint {https://arxiv.org/abs/1210.8342} {arXiv:1210.8342} \BibitemShut {NoStop}%
\bibitem [{\citenamefont {Mancini}\ and\ \citenamefont {Tombesi}(1994)}]{Mancini1994a}%
  \BibitemOpen
  \bibfield  {author} {\bibinfo {author} {\bibfnamefont {S.}~\bibnamefont {Mancini}}\ and\ \bibinfo {author} {\bibfnamefont {P.}~\bibnamefont {Tombesi}},\ }\bibfield  {title} {\bibinfo {title} {{Quantum noise reduction by radiation pressure}},\ }\href {https://doi.org/10.1103/PhysRevA.49.4055} {\bibfield  {journal} {\bibinfo  {journal} {Physical Review A}\ }\textbf {\bibinfo {volume} {49}},\ \bibinfo {pages} {4055} (\bibinfo {year} {1994})}\BibitemShut {NoStop}%
\bibitem [{\citenamefont {Fabre}\ \emph {et~al.}(1994)\citenamefont {Fabre}, \citenamefont {Pinard}, \citenamefont {Bourzeix}, \citenamefont {Heidmann}, \citenamefont {Giacobino},\ and\ \citenamefont {Reynaud}}]{Fabre1994}%
  \BibitemOpen
  \bibfield  {author} {\bibinfo {author} {\bibfnamefont {C.}~\bibnamefont {Fabre}}, \bibinfo {author} {\bibfnamefont {M.}~\bibnamefont {Pinard}}, \bibinfo {author} {\bibfnamefont {S.}~\bibnamefont {Bourzeix}}, \bibinfo {author} {\bibfnamefont {A.}~\bibnamefont {Heidmann}}, \bibinfo {author} {\bibfnamefont {E.}~\bibnamefont {Giacobino}},\ and\ \bibinfo {author} {\bibfnamefont {S.}~\bibnamefont {Reynaud}},\ }\bibfield  {title} {\bibinfo {title} {{Quantum-noise reduction using a cavity with a movable mirror}},\ }\href {https://doi.org/10.1103/PhysRevA.49.1337} {\bibfield  {journal} {\bibinfo  {journal} {Physical Review A}\ }\textbf {\bibinfo {volume} {49}},\ \bibinfo {pages} {1337} (\bibinfo {year} {1994})}\BibitemShut {NoStop}%
\bibitem [{\citenamefont {Aspelmeyer}\ \emph {et~al.}(2014)\citenamefont {Aspelmeyer}, \citenamefont {Kippenberg},\ and\ \citenamefont {Marquardt}}]{Aspelmeyer2014}%
  \BibitemOpen
  \bibfield  {author} {\bibinfo {author} {\bibfnamefont {M.}~\bibnamefont {Aspelmeyer}}, \bibinfo {author} {\bibfnamefont {T.~J.}\ \bibnamefont {Kippenberg}},\ and\ \bibinfo {author} {\bibfnamefont {F.}~\bibnamefont {Marquardt}},\ }\bibfield  {title} {\bibinfo {title} {{Cavity optomechanics}},\ }\href {https://doi.org/10.1103/RevModPhys.86.1391} {\bibfield  {journal} {\bibinfo  {journal} {Reviews of Modern Physics}\ }\textbf {\bibinfo {volume} {86}},\ \bibinfo {pages} {1391} (\bibinfo {year} {2014})},\ \Eprint {https://arxiv.org/abs/1303.0733} {arXiv:1303.0733} \BibitemShut {NoStop}%
\bibitem [{\citenamefont {Bensemhoun}\ \emph {et~al.}(2024)\citenamefont {Bensemhoun}, \citenamefont {Gonzalez-Arciniegas}, \citenamefont {Pfister}, \citenamefont {Labont{\'e}}, \citenamefont {Etesse}, \citenamefont {Martin}, \citenamefont {Tanzilli}, \citenamefont {Patera},\ and\ \citenamefont {D'Auria}}]{Bensemhoun2024}%
  \BibitemOpen
  \bibfield  {author} {\bibinfo {author} {\bibfnamefont {A.}~\bibnamefont {Bensemhoun}}, \bibinfo {author} {\bibfnamefont {C.}~\bibnamefont {Gonzalez-Arciniegas}}, \bibinfo {author} {\bibfnamefont {O.}~\bibnamefont {Pfister}}, \bibinfo {author} {\bibfnamefont {L.}~\bibnamefont {Labont{\'e}}}, \bibinfo {author} {\bibfnamefont {J.}~\bibnamefont {Etesse}}, \bibinfo {author} {\bibfnamefont {A.}~\bibnamefont {Martin}}, \bibinfo {author} {\bibfnamefont {S.}~\bibnamefont {Tanzilli}}, \bibinfo {author} {\bibfnamefont {G.}~\bibnamefont {Patera}},\ and\ \bibinfo {author} {\bibfnamefont {V.}~\bibnamefont {D'Auria}},\ }\bibfield  {title} {\bibinfo {title} {Multipartite entanglement in bright frequency combs out of microresonators},\ }\href {https://doi.org/https://doi.org/10.1016/j.physleta.2023.129272} {\bibfield  {journal} {\bibinfo  {journal} {Physics Letters A}\ }\textbf {\bibinfo {volume} {493}},\ \bibinfo {pages} {129272} (\bibinfo {year} {2024})}\BibitemShut {NoStop}%
\bibitem [{\citenamefont {Zhang}\ \emph {et~al.}(2021)\citenamefont {Zhang}, \citenamefont {Menotti}, \citenamefont {Tan}, \citenamefont {Vaidya}, \citenamefont {Mahler}, \citenamefont {Helt}, \citenamefont {Zatti}, \citenamefont {Liscidini}, \citenamefont {Morrison},\ and\ \citenamefont {Vernon}}]{Zhang2021}%
  \BibitemOpen
  \bibfield  {author} {\bibinfo {author} {\bibfnamefont {Y.}~\bibnamefont {Zhang}}, \bibinfo {author} {\bibfnamefont {M.}~\bibnamefont {Menotti}}, \bibinfo {author} {\bibfnamefont {K.}~\bibnamefont {Tan}}, \bibinfo {author} {\bibfnamefont {V.~D.}\ \bibnamefont {Vaidya}}, \bibinfo {author} {\bibfnamefont {D.~H.}\ \bibnamefont {Mahler}}, \bibinfo {author} {\bibfnamefont {L.~G.}\ \bibnamefont {Helt}}, \bibinfo {author} {\bibfnamefont {L.}~\bibnamefont {Zatti}}, \bibinfo {author} {\bibfnamefont {M.}~\bibnamefont {Liscidini}}, \bibinfo {author} {\bibfnamefont {B.}~\bibnamefont {Morrison}},\ and\ \bibinfo {author} {\bibfnamefont {Z.}~\bibnamefont {Vernon}},\ }\bibfield  {title} {\bibinfo {title} {Squeezed light from a nanophotonic molecule},\ }\href {https://doi.org/10.1038/s41467-021-22540-2} {\bibfield  {journal} {\bibinfo  {journal} {Nature Communications}\ }\textbf {\bibinfo {volume} {12}},\ \bibinfo {pages} {2233} (\bibinfo {year} {2021})}\BibitemShut {NoStop}%
\end{thebibliography}%

\end{document}